\documentclass[twocolumn]{aastex701}

\usepackage{xcolor,tikz}
\usepackage{newtxtext,newtxmath,soul}
\usepackage{multirow}
\usepackage[T1]{fontenc}
\usepackage{gensymb}
\usepackage{tablefootnote}
\usepackage{makecell}
\usepackage{comment}
\usepackage{amsmath}
\usepackage{graphicx} 
\makeatletter
\renewcommand{\thepage}{\ifnum\value{page}=22 \else\arabic{page}\fi}
\makeatother

\begin{document}

\title{A Comprehensive Study of the Magnetic White Dwarfs in DESI DR1}

\author[orcid=0000-0001-7143-0890]{Adam Moss}
\affiliation{Department of Astronomy, University of Florida, Bryant Space Science Center, Stadium Road, Gainesville, FL 32611, USA}
\email[show]{adamgregorymoss@gmail.com}  
\author[0000-0003-2368-345X]{Pierre Bergeron}
\affiliation{Département de Physique, Université de Montréal, C.P. 6128, Succ. Centre-Ville, Montréal, Québec H3C 3J7, Canada}
\email{bergeron@astro.umontreal.ca}
\author[0000-0001-6098-2235]{Mukremin Kilic}
\affiliation{Homer L. Dodge Department of Physics and Astronomy, University of Oklahoma, 440 W. Brooks St., Norman, OK 73019, USA}
\email{kilic@ou.edu}
\author[0000-0002-4462-2341]{Warren R. Brown}
\affiliation{Center for Astrophysics | Harvard \& Smithsonian,
60 Garden Street, Cambridge, MA 02138, USA}
\email{wbrown@cfa.harvard.edu}
\author[0000-0002-4818-7885]{Jamie Tayar}
\affiliation{Department of Astronomy, University of Florida, Bryant Space Science Center, Stadium Road, Gainesville, FL 32611, USA}
\email{jtayar@ufl.edu}

\begin{abstract}

As the end point of stellar evolution for most stars, white dwarfs offer powerful insight into stellar magnetism. Multiple formation channels are required to explain the observed properties of magnetic white dwarfs, and the mechanisms that produce unusual objects have potential to unlock new realms of physics. The large samples made possible by spectroscopic surveys enable study of these processes on an unprecedented level. Here we perform a detailed model atmosphere analysis of the magnetic white dwarfs in DESI Data Release 1. We identify 831 magnetic systems, 760 of which we fit using an offset dipole model. Of these, 126 targets require more in-depth analysis as they possess shallower or sharper absorption features than predicted by the offset dipole model. For these we use a patchy offset dipole atmosphere model in which the magnetic field only contributes to the line opacities at certain regions of the visible stellar disk. We compare the observed distribution of magnetic white dwarfs to the mechanisms that generate magnetic fields. We find that a core-convective dynamo that forms on the main-sequence matches the parameters of many of the magnetic systems, while a dynamo that fills the radiative interior on the giant branch reproduces many of the lower mass objects that the core-convective dynamo cannot. This sample is the largest single data release of magnetic white dwarfs to date. The diversity of formation channels and the frequency of objects with unusual field geometries highlight the power of multiplexed surveys to advance our knowledge of stellar magnetism. 

\end{abstract}

\keywords{\uat{White dwarf stars}{1799} --- \uat{Stellar magnetic fields}{1710} --- \uat{Stellar evolution}{1599} --- \uat{Compact objects}{288}}

\section{Introduction}

Magnetism impacts stellar evolution in a variety of ways, such as slowing down surface rotation via magnetic braking \citep{Weber67} to redistributing angular momentum in the stellar interior \citep{Mestel87}. Thus a complete understanding of how stars evolve must incorporate the effects of magnetic fields, from the star's birth all the way to its death. This includes white dwarf remnants, which serve as an important anchor given that any predictions of field evolution must match what we observe on the white dwarf sequence. A large sample of magnetic white dwarfs (MWDs) therefore provides insight not just into white dwarf physics but also the evolution of their progenitor stars. 

MWDs are fairly common among the white dwarf population. \citet{Bagnulo21} conducted polarimetric observations of the white dwarfs within the local 20 pc volume and found that 22\% of all white dwarfs are magnetic. With a field sensitivity of 1 kG for WDs with hydrogen-dominated atmospheres (DAs), extremely weak fields can be detected via polarimetry to provide an unbiased view of the MWD population. Unfortunately, this method requires high signal-to-noise (see \citealt{Landstreet15} for a summary of the diagnostic content of spectropolarimetry) which clashes with the intrinsic faintness of white dwarfs. Hence this technique of studying MWDs is typically limited to objects within a few tens of parsecs from the Sun. 

Alternatively, magnetic fields in white dwarfs are often strong enough to produce Zeeman splitting in absorption lines, which can be identified with spectroscopy. While this method is limited to field strengths of $\sim$0.1 MG and higher \citep{Kepler13,Ferrario15,Kawka20,Moss25b}, as well as white dwarfs that are warm enough to show absorption features, it can be applied to a much larger sample of objects given the less restrictive brightness requirements. In particular, the use of wide-field spectroscopic surveys allows for the spectral classification of hundreds of thousands of white dwarfs, and for identifying rare objects such as variable systems via multi-epoch spectra. Wide-field surveys are therefore a powerful tool for understanding multiple aspects of white dwarf magnetism, such as the origin of their fields and how the fields impact evolutionary processes. 

Historically, the Sloan Digital Sky Survey (SDSS) served as the primary driver of MWD identification, with hundreds of objects identified throughout its life cycle \citep{Gansicke02,Schmidt03,Vanlandingham05,Kepler13}. The number of MWD detections from SDSS has helped shape our understanding of stellar magnetism in numerous ways. \citet{Kulebi09} analyzed 97 MWDs with hydrogen-rich atmospheres (DAHs) and found that Ap/Bp stars could not replicate the frequency or field strengths of the MWD sample, suggesting an alternative formation mechanism. \citet{Amorim23} determined the field strength of 804 DAHs and found that the field strength and magnetic fraction increase with decreasing $T_{\rm eff}$. \citet{Hardy23a,Hardy23b} similarly analyzed 651 DAHs and 79 MWDs with helium-rich atmospheres respectively and found that many spectral features could not be reproduced with models, indicating a more complex field structure or chemical composition. \citet{Moss25b} analyzed all MWDs in the 100 pc volume in the SDSS footprint and found evidence of two distinct classes of MWDs that come from separate evolution channels, similar to previous findings from volume-limited polarimetric surveys \citep{Bagnulo21,Bagnulo22}.

The SDSS has considerably expanded our knowledge of MWDs, and there are several other ongoing spectroscopic surveys that will have similar impact. SDSS-V \citep{Kollmeier19}, the Dark Energy Spectroscopic Instrument (DESI, \citealt{DESI25}), the 4-metre Multi-Object Spectroscopic Telescope (4MOST, \citealt{DeJong19}), and the William Herschel Telescope Enhanced Area Velocity Explorer (WEAVE, \citealt{Jin24}) will obtain spectra on thousands of high-confidence white dwarf candidates from Gaia DR3 \citep{GaiaDR321,Fusillo21}. Many of these will be magnetic, which will lead to new insights into the MWD population and allow us to test new modeling techniques to reproduce the observed spectra. \citet{Manser24} identified 2706 white dwarfs in the DESI Early Data Release, 56 of which were classified as magnetic. 70\% of these were new discoveries, demonstrating the power of these spectroscopic surveys to find new MWDs. With larger samples, we can resolve white dwarf magnetism at a level previously out of reach, allowing us to populate sparsely sampled regions of parameter space, test new formation and evolution models, and uncover rare objects.

Here we present our analysis on the entire MWD sample in DESI Data Release 1. This paper follows the analysis of the DESI DR1 white dwarf sample by \citet{Kilic26hot,Kilic26cold} where they analyzed all white dwarfs in DESI DR1 with $G_{BP}-G_{RP} \leq0$ and $G_{BP}-G_{RP} > 0$ respectively. We discuss the sample selection in Section 2, followed by a discussion of our fitting method in Section 3. We then discuss the sample properties, namely the mass, temperature, and field strength distributions in Section 4. Finally we conclude in Section 5. 

\section{Sample Selection}

DESI, mounted on the 4m Mayall telescope at the Kitt Peak National Observatory, is a multi-object spectrograph capable of obtaining spectra on $\sim$5000 objects per pointing \citep{DESI22}. The instrument covers a wavelength range of 3600$-$9800 \AA\ at a full width at half maximum resolution of $\sim$1.8 \AA. DESI DR1 covers the observing dates 2021 May through 2022 June, with 66,811 potential white dwarf candidates in the observing footprint as part of the BRIGHT program \citep{Cooper23}. 

As described in \citet{Kilic26hot,Kilic26cold}, there are 47,795 white dwarf candidates in common between DESI DR1 and the Gaia DR3 white dwarf catalog from \citet{Fusillo21}. We determine each object's spectral type via visual inspection, and we identify magnetic targets based on Zeeman splitting of optical absorption lines. Since a high magnetic field strength ($\sim$100 MG) can smear out absorption lines, we can detect fields if a target has a featureless spectrum and if it is at a sufficiently high effective temperature such that we would expect to see H or He lines. For targets with $G_{BP}-G_{RP} \leq0$, which corresponds to $T_{\rm eff} \approx 10,000$ K, both H and He lines are visible such that our classification and analysis is highly robust above this cutoff. White dwarfs with He atmospheres produce featureless spectra (DCs) below this limit, so we do not label these as magnetic. While it is possible some possess fields, they cannot be confirmed at this time. Additionally, we are biased against objects with weak fields ($B < 1$ MG) as the splitting is too weak to detect in medium-resolution spectra. 

Our visual inspection yields 831 MWDs. Since we are solely interested in the single white dwarf population, we remove all objects from the non-magnetic sample that were not classified in \citet{Kilic26hot,Kilic26cold} as single white dwarfs. This includes subdwarfs, AM CVn, CVs, double white dwarf binaries, white dwarf + M dwarf binaries, etc. This leaves 42,441 non-magnetic objects, yielding a magnetic fraction of 1.9\%. Similarly, \citet{Amorim26b} identified 765 MWDs out of 40,439 single white dwarfs (1.9\%) while \citet{Swan26} identified 688 MWDs out of 43,215 single objects (1.6\%).

This is a notably lower fraction than other spectroscopic surveys such as \citet{Kepler13}, who found a magnetic fraction of 4\% among DAs in SDSS DR7, and \citet{Moss25b}, who obtained a fraction of 5.2\% across all spectral types in the SDSS 100 pc sample. This is likely due to the low S/N of the DESI spectra, which makes identifying Zeeman-split features challenging. The median S/N of the magnetic sample is 10.3, and the median for the non-magnetic sample is only 8.0. Hence the 1.9\% magnetic fraction is likely a lower limit for this sample. Of our 831 MWDs, 790 (95\%) have hydrogen-dominated atmospheres, comparable to that of the SDSS 100 pc sample (91\%, \citealt{Moss25b}).

\begin{figure}[!ht]
    \centering
    \includegraphics[width=3.75in, clip=true, trim=0.2in 0in 0in 0in]{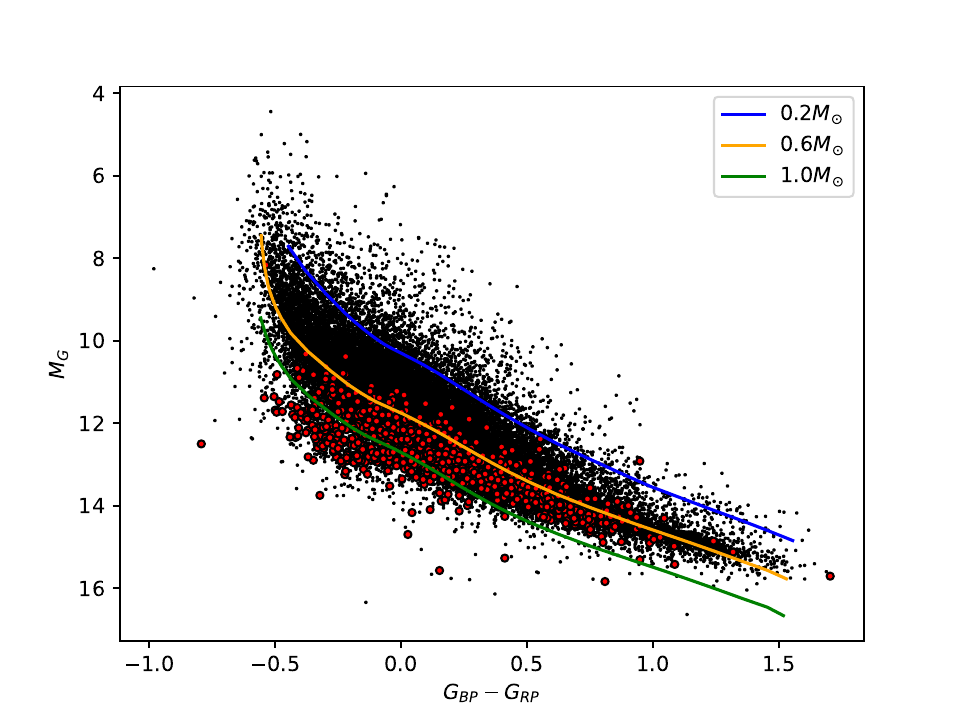}
    \caption{Gaia color-magnitude diagram of single white dwarfs in DESI DR1 (black points) with the MWDs marked by red points. The colored lines are the cooling tracks for a $0.2 M_{\odot}$, $0.6 M_{\odot}$, and $1.0 M_{\odot}$ white dwarf with pure H atmospheres.}
    \label{fig1} 
\end{figure}

Figure \ref{fig1} shows the Gaia color-magnitude diagram for our DESI sample of single white dwarfs with the MWDs highlighted and the evolutionary sequences for a $0.2 M_{\odot}$, $0.6 M_{\odot}$, and $1.0 M_{\odot}$ white dwarf with pure H atmospheres \citep{Tremblay11,Blouin19,Bedard20}. We immediately see that magnetism is quite rare among lower mass objects, with the vast majority of MWDs having masses higher than 0.6 $M_{\odot}$.

\section{Model Atmosphere Analysis}
\subsection{Magnetic Fitting Procedure}

Our modeling process is discussed in detail in \citet{Kilic26hot,Kilic26cold}. Briefly, we use the photometric technique described by \citet{Bergeron19} to constrain the effective temperature and solid angle, thus the radius and mass. For each object we calculate the specific intensities at the surface, $I(\nu,\mu,\tau_{\nu}=0)$, which are obtained by solving the radiative transfer equation for various field strengths and values of $\mu$ ($\mu= \cos \theta$, where $\theta$ is the angle between the propagation of light and the normal to the surface of the star). The line displacements and oscillator strengths of the Zeeman components of H$\alpha$ through H$\delta$ are kindly provided by S. Jordan (see also \citealt{Hardy23a,Hardy23b,Moss24}). The total line opacity is calculated as the sum of the individual Stark-broadened Zeeman components and is normalized to that resulting from the zero-field solution. The emergent Eddington flux $H_\nu$ is then calculated on
the fly during the fitting process by numerically integrating the specific intensity over the visible surface of the disc.

We use the offset dipole model \citep{Martin84,Achilleos89} to determine the magnetic parameters. This geometry assumes a field structure generated by a dipole, and then the dipole is offset along the z-axis (see Figure 1 of \citealt{Achilleos89} for a diagram of the geometry). The model contains three parameters that produce the spectrum: the dipole field strength $B_d$, the viewing angle $i$ between the line of sight and the axis of symmetry, and the dipole offset $a_z$ measured in units of stellar radii (see Figure 4 of \citealt{Bergeron92} for example synthetic spectra). We use the \texttt{PIKAIA} genetic algorithm \citep{Charbonneau95} to determine the best-fitting model. 

\begin{figure}[!ht]
    \centering
    \includegraphics[width=3.5in, clip=true, trim=0in 3in 0in 2in]{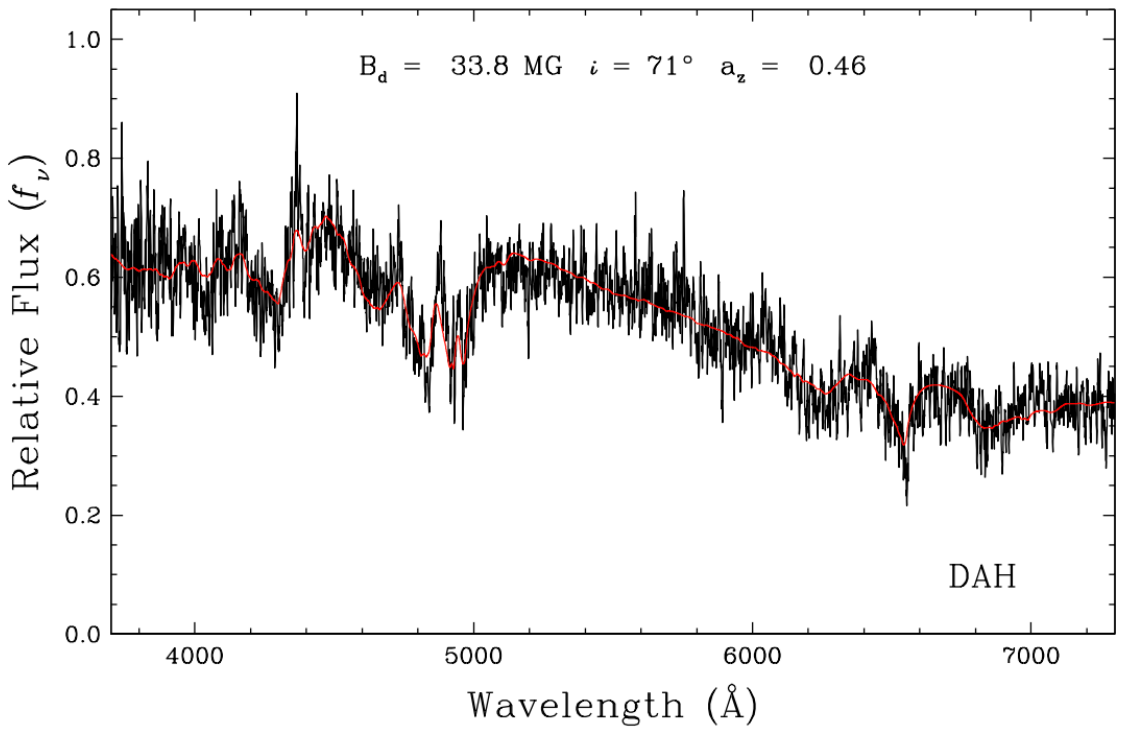}
    \includegraphics[width=3.5in, clip=true, trim=0in 3in 0in 2in]{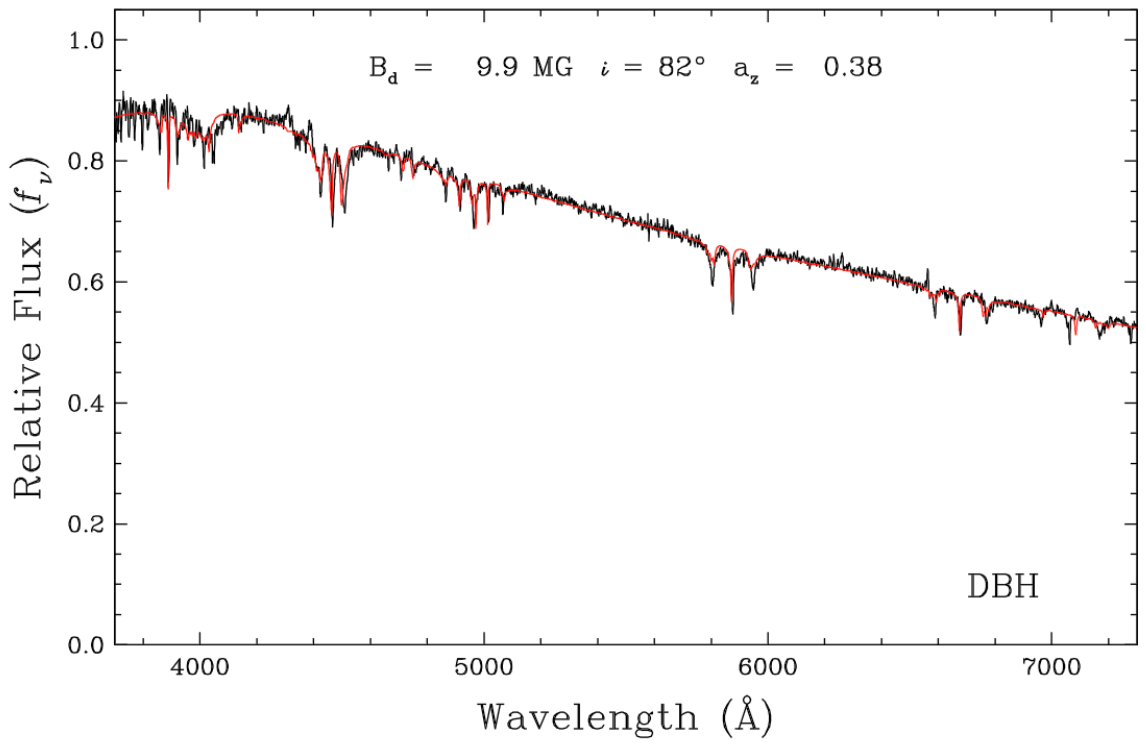}
    \caption{Offset dipole fits to WDJ011423.35+160727.57 (top panel) and WDJ001742.44+004137.35 (bottom panel).}
    \label{fig2} 
\end{figure}

\begin{figure}[!ht]
    \centering
    \includegraphics[width=3.5in, clip=true, trim=0in 1in 0in 1.45in]{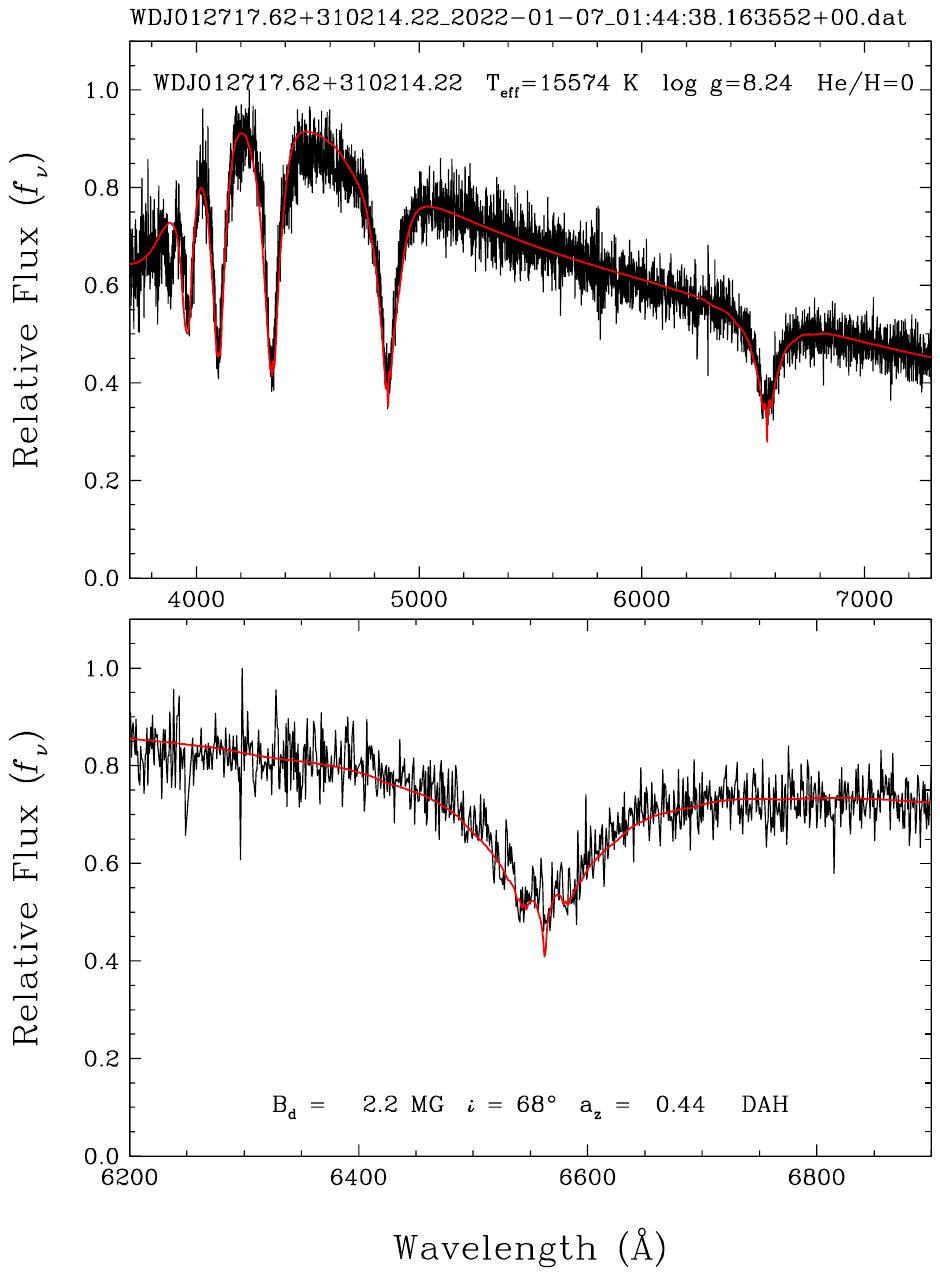}
    \caption{Offset dipole fit to a DAH with a low field strength ($B=2.2$ MG). Here we fit the H$\alpha$ region (bottom panel) and then refit the entire spectrum while holding the magnetic parameters fixed (top panel).}
    \label{fig3} 
\end{figure}

The offset dipole model yields excellent fits to many of our objects which do not require further analysis. Figure \ref{fig2} shows fits to a DAH and DBH from our sample, WDJ011423.35+160727.57 and WDJ001742.44+004137.35 respectively. We successfully reproduce even weak features visible in these spectra. Some targets possess very weak fields that require higher wavelength resolution in the models to reproduce the subtle splitting observed in the line cores. For these objects, we increase the wavelength resolution and use only the H$\alpha$ region to determine the best fitting magnetic parameters. We then refit the entire spectrum while holding these magnetic parameters fixed. An example object with a weak field is shown in Figure \ref{fig3}. The low field strength of 2.2 MG yields splitting in the H$\alpha$ line of only $\sim$10s of Angstroms, which we successfully reproduce in our model fit.

Our analysis of the DESI sample was initially split into two categories: hot WDs that are bluer than $G_{BP}-G_{RP} =0$, and cold WDs that are redder than this limit. In the hot sample, \citet{Kilic26hot} identified 296 MWDs. Since this sample consists of hot white dwarfs, we can identify magnetic fields in featureless spectra, as a strong field will smear out the absorption profiles that would otherwise be present. Since we are unable to constrain the atmosphere composition in those specific objects, we label such objects as ``DAH?''. We label objects for which a magnetic field detection is tentative as ``DAH:''. We obtain magnetic fits for all of the hydrogen and helium-dominated atmosphere objects except for a DABH, WDJ084716.22+484220.32. This known variable system possesses an inhomogeneous distribution of H and He in its atmosphere \citep{Moss25a}, but the spectrum does not show clear evidence of magnetism besides the shallow absorption features. Hence we do not attempt a magnetic fit. This leaves 288 hot MWDs with magnetic fits.

For the cold sample, \citet{Kilic26cold} identified 535 MWDs. Of these, we obtain fits for all of the DAH, DAH?, and DAH: objects as well as two MWDs with hydrogen emission features (DAHe). While there are 37 objects classified as DAHe in the sample, only two show absorption features that can be fit with our model. Hence we have 472 cold MWDs with fits that we use in our analysis, for 760 total. 

\begin{deluxetable}{ccc}
\tabletypesize{\footnotesize}
\tablecolumns{3} \tablewidth{0pt}
\tablecaption{Spectral types of the MWDs in our DESI DR1 analysis.} \label{tab1}
\tablehead{\colhead{Spectral Type} & \colhead{Hot Sample} & \colhead{Cold Sample}}
\startdata
DAH & 206 & 389 \\
DAH? & 69 & 29 \\
DAH: & 2 & 52 \\
DAHe & 0 & 36 \\
DAHe? & 0 & 1 \\
DBH & 6 & 0 \\
DBH? & 3 & 0 \\
DABH & 2 & 0 \\
DBAH & 1 & 0 \\
DQH & 5 & 6 \\
DQH? & 2 & 1 \\
DQH: & 0 & 3 \\
DZH & 0 & 15 \\
DZH: & 0 & 1 \\
DAZH & 0 & 2 \\
\enddata
\end{deluxetable}

We have identified a noticeably larger amount of MWDs (831) than \citet{Swan26} (688). The discrepancy can be explained by our larger DAH sample (594 compared to 522) and our inclusion of tentative detections (58). Specifically, we classify 67 objects as magnetic that \citet{Swan26} classify as DC white dwarfs. Many of these are hot and thus should show absorption features but do not, so we assume that they have strong fields even if we cannot constrain the atmospheric composition.

We also find discrepancies between magnetic objects and those with carbon-rich and metal-rich atmospheres. We find nine more DZH and ten more DQH white dwarfs. Additionally, we find four objects that we classify as DAH or DAH? while \citet{Swan26} classify them as DQ or DAQ. We reclassified two of these, J0204$-$0137 and J1750+6642, as a DQ and DAQ respectively after comparing our classifications and redoing the fits with a carbon-rich atmosphere model. However, the other two, J1259+0823 and J1741+7638, cannot be reproduced with a carbon-rich model; only a magnetic model reproduces their spectra. \citet{Swan26} only attempted model fits assuming a pure H, pure He, or mixed H+He atmosphere. Hence these two objects are indeed magnetic.

\subsection{Patchy Atmospheres}

\subsubsection{An Alternative Model}

The offset dipole model we have discussed so far is effectively a homogeneous model: the entire surface is assumed to be H or He-rich depending on the spectral type, and thus the whole visible stellar disk contributes to the corresponding line opacities in the spectrum. We identify 126 objects which we cannot fit with this homogeneous offset dipole model. These targets typically have sharper or weaker absorption features than what our standard model predicts, which results in poor fits. All 126 objects that require patchy fits show H lines in their spectra. Hence these are not cases where a strong field has wiped out the line profiles like in the objects labeled as ``DAH?''.

For these objects, we use an inhomogeneous model, which is an extension of the homogeneous offset dipole model and uses the same magnetic geometry. This patchy model assumes H polar caps and a He belt, and we introduce a new parameter, $\theta_c$, which defines the size of the polar caps. For $\theta_c=90\degree$, the entire surface is H (in the case of DAHs) and thus the patchy model reduces to the homogeneous offset dipole model. The goal of this model is to restrict the regions of the visible surface that contribute to the line opacities. In this way, we are effectively creating a proxy for field complexity that is not captured by the homogeneous offset dipole model. As we will show, this method is highly successful in yielding excellent fits to the spectra. 

We stress that these results do not represent a unique reconstruction of the surface geometry, but they are instead proxies for a complex field and chemical structure. Of the 126 targets that require patchy fits, 66 of them are below 11,000 K such that we would not expect to see He lines in their spectra. Of the remaining 61 that are hot enough to show He features, only three show clear evidence of them. Hence the targets that require patchy fits could simply have complex field structures that deviate significantly from the homogeneous model. There exist degeneracies between $B_d$, $i$, $a_z$, and $\theta_c$ such that other geometries could be considered to obtain similar quality fits. It is also beyond the scope of this paper to create a map of every object and determine what the true geometry is in each case. While a model that includes rotational smearing and a quadrupole could provide a more realistic interpretation of the field structure, development and implementation of that model would need to be addressed in the future. Instead, the goal is to simply identify and present cases where the homogeneous offset dipole geometry fails, and to apply the patchy model to improve the fit quality and provide an initial estimate of the field strength. 

\subsubsection{Example Fits}

\begin{figure}[!ht]
    \centering
    \includegraphics[width=3.5in, clip=true, trim=0.7in 1in 0.7in 5.6in]{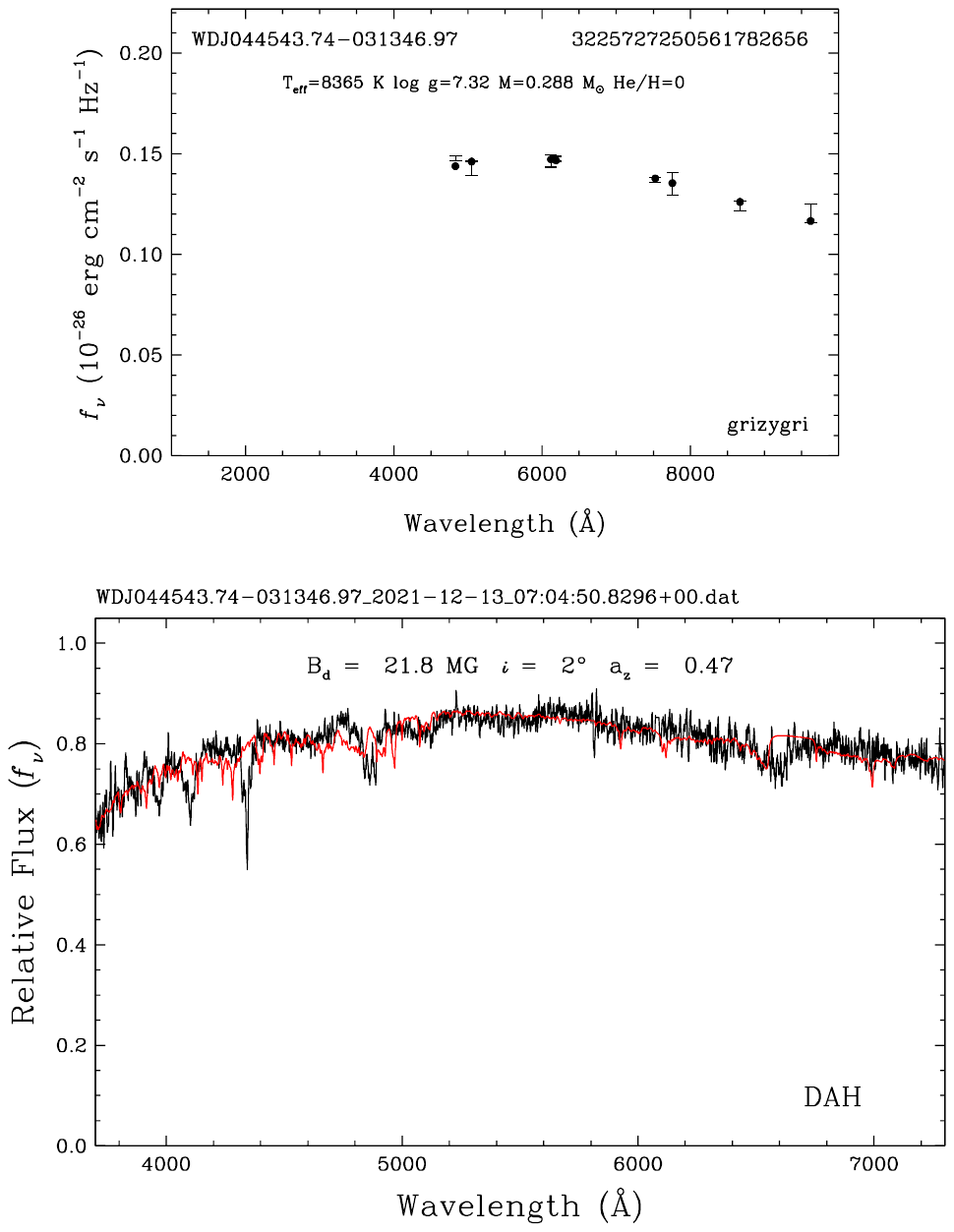}
    \includegraphics[width=3.5in, clip=true, trim=0.9in 1.2in 1in 1.45in]{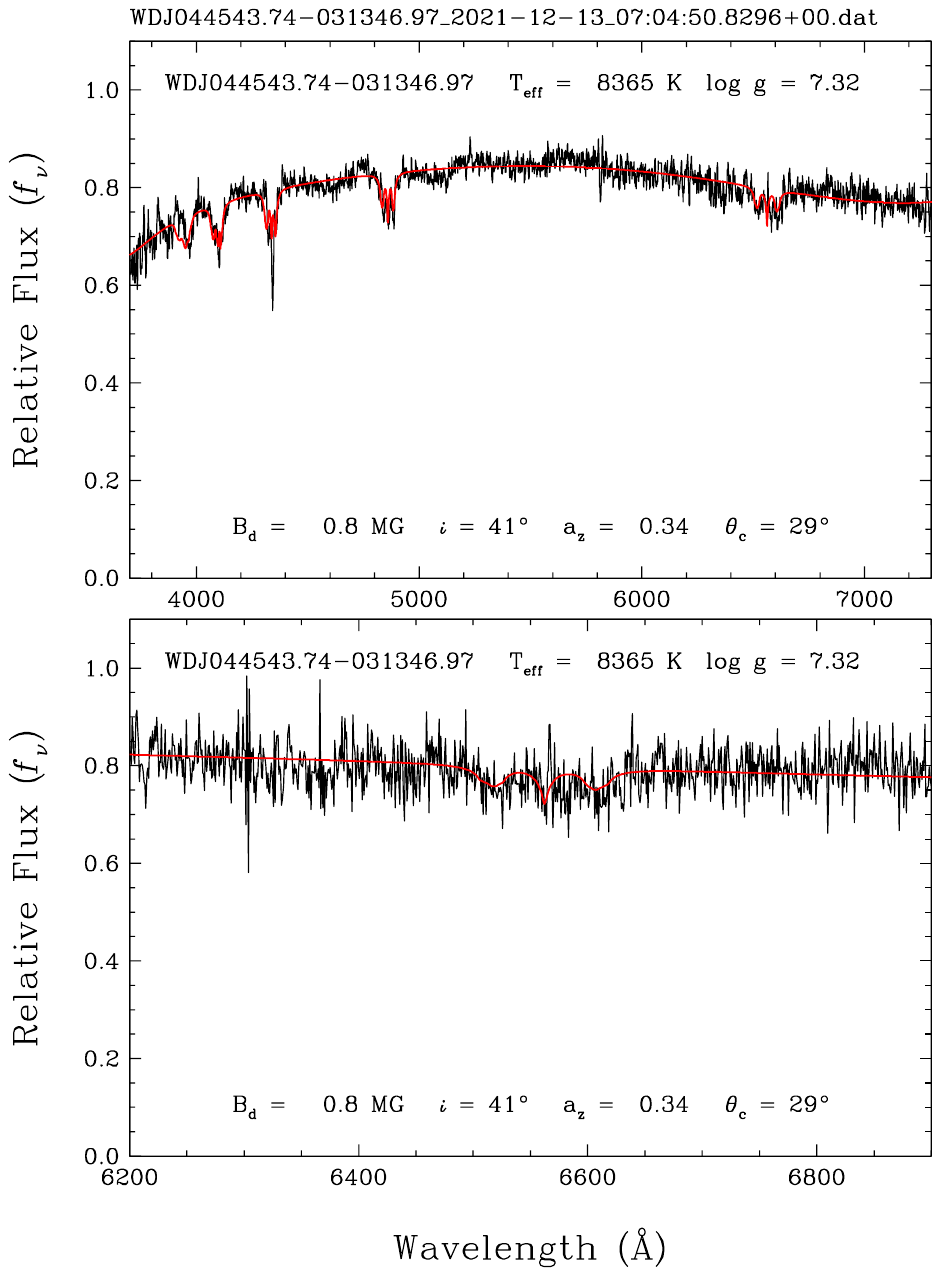}
    \caption{Top: Homogeneous fit to a cool DAH, WDJ044543.74$-$031346.97. The fit fails to reproduce the weak but noticeable Balmer lines. Middle: Patchy fit to the entire spectrum using the parameters obtained from the fit to the H$\alpha$ region. This fit reproduces all of the features, with a much weaker field strength. Bottom: Patchy fit to the H$\alpha$ region. The patchy model produces an excellent fit to H$\alpha$ with a more realistic field strength.}
    \label{figj0445} 
\end{figure}

Figure \ref{figj0445} shows our homogeneous and patchy fits to an example cool DAH, WDJ044543.74$-$031346.97. Here the Balmer lines are quite weak compared to what would be expected of a DA at $T_{\rm eff} = 8365$ K. The splitting of the Zeeman components is similarly quite weak, much weaker than what one would expect from the homogeneous  solution of 21.8 MG. Hence this target does not have a strong field, which would greatly shift the positions of the Balmer lines and leave no trace of weak splitting. Since He cannot be detected below 11,000 K, any regions on the magnetic surface that contain He would not contribute to the line opacities in the spectrum. As a result, the observed spectral features would resemble a field structure that is restricted to only specific regions of the surface rather than one that extends over the entire visible stellar disk. Hence our patchy model for these cool objects introduces chemical inhomogeneities by using H caps and a He belt, which accomplishes two goals: it reproduces the weakened H lines by restricting the fraction of the visible stellar disk that is covered by H, and consequently it restricts the magnetic field distribution that contributes to the observed absorption profiles. The resulting fit is excellent as shown in Figure \ref{figj0445}. 

\begin{figure*}
    \centering
    \includegraphics[width=7in, clip=true, trim=0in 0in 0in 0in]{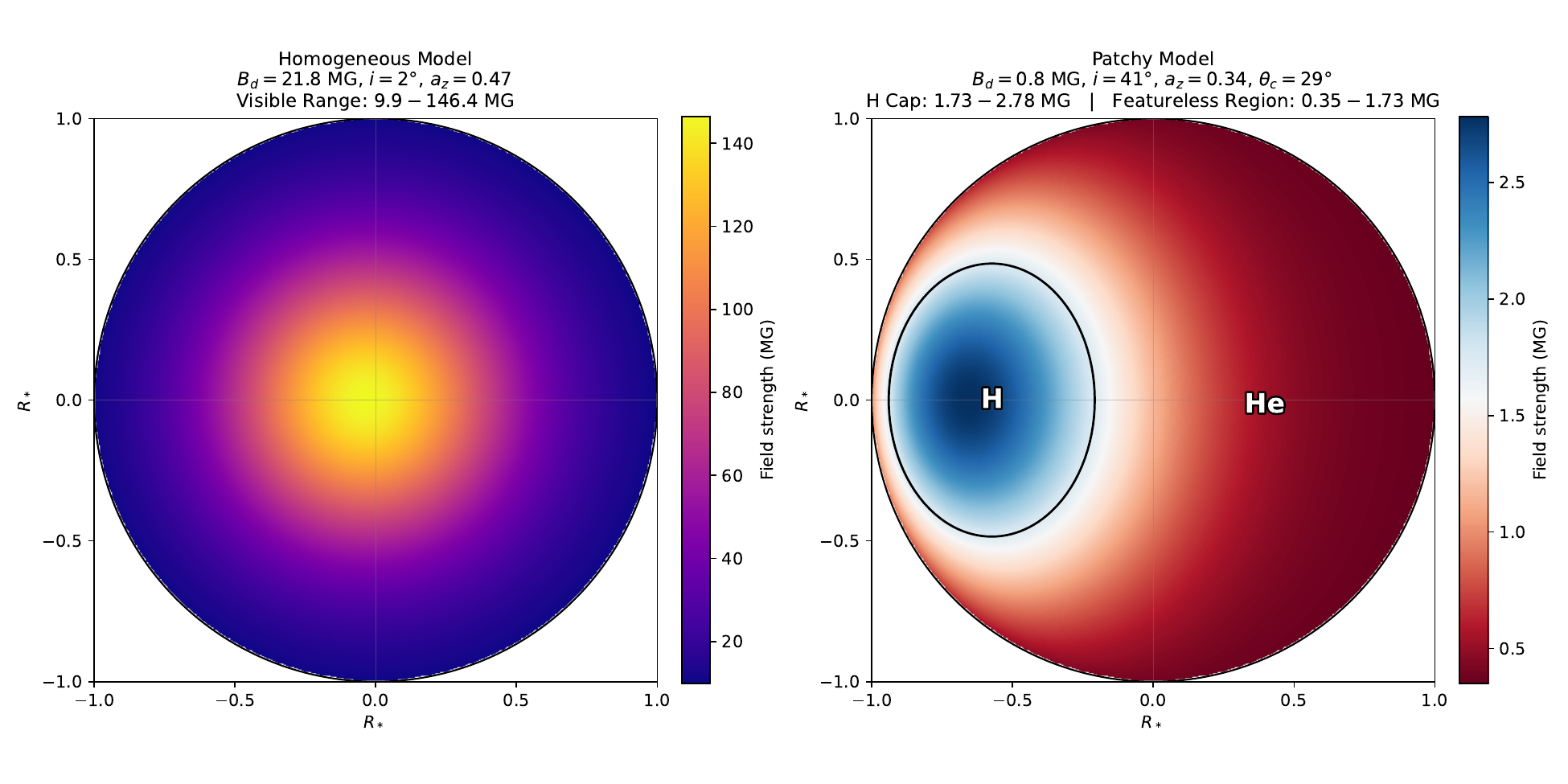}
    \caption{Topology map of WDJ044543.74$-$031346.97 using our best-fit homogeneous H solution (left panel) and our patchy atmosphere solution (right panel). The homogeneous model yields a pole-on viewing angle and a wide range of field strengths across the visible stellar disk. The patchy solution yields a much narrower range of field strengths, with the H polar cap contributing to the line opacities while the He region does not.}
    \label{figj0445map}
\end{figure*}

To better illustrate the differences between the homogeneous and patchy fit, Figure \ref{figj0445map} shows a topology map for the best solution of each model. In the homogeneous case, the viewer looks nearly pole-on at the surface, and the local field strengths span a wide range over the visible disk from 9.9 to 146.4 MG. It is worth noting that the dipole field strength of 21.8 MG is significantly lower than the maximum local field in this calculation. The field strength from our fit is not the average field across the stellar disk, nor is it the intrinsic field strength of the object. It is only a proxy that reproduces the observed field distribution on the surface. Observing the same target at different phases, or observing multiple phases in one long exposure, will likely produce different $B_d$ values. 

This is even more apparent in the case of the patchy model. In contrast to the homogeneous model, the patchy model produces a much lower range of surface field strengths. The viewing angle is now pointed between the pole and the equator, and only the $29\degree$ polar cap contributes to the line opacities. An important result here is that the field strengths within the cap, 1.73$-$2.78 MG, are larger than the best-fit field strength of 0.8 MG, even though only the cap contributes to the line intensities. The dipole field strength should thus not be interpreted as what we measure from the H features. Rather, it is a characterization of the offset dipole geometry. 

The failure of the homogeneous model here, specifically the converged high dipole field strength and large range of local field strengths despite the observed weak splitting, can be explained by the wide surface coverage. Since the homogeneous model is forced to have H over the entire surface, the fitting routine uses a broad distribution of magnetic field strengths to spread the Zeeman opacity over a large wavelength range and thereby weaken the Balmer features. Once the patchy model can reduce the fraction of the surface contributing H opacity, it no longer needs to use the magnetic geometry to artificially weaken the lines, and it converges to the much more reasonable field distribution in the H cap.

\begin{figure}[!ht]
    \centering
    \includegraphics[width=3.5in, clip=true, trim=0in 3in 0in 3in]{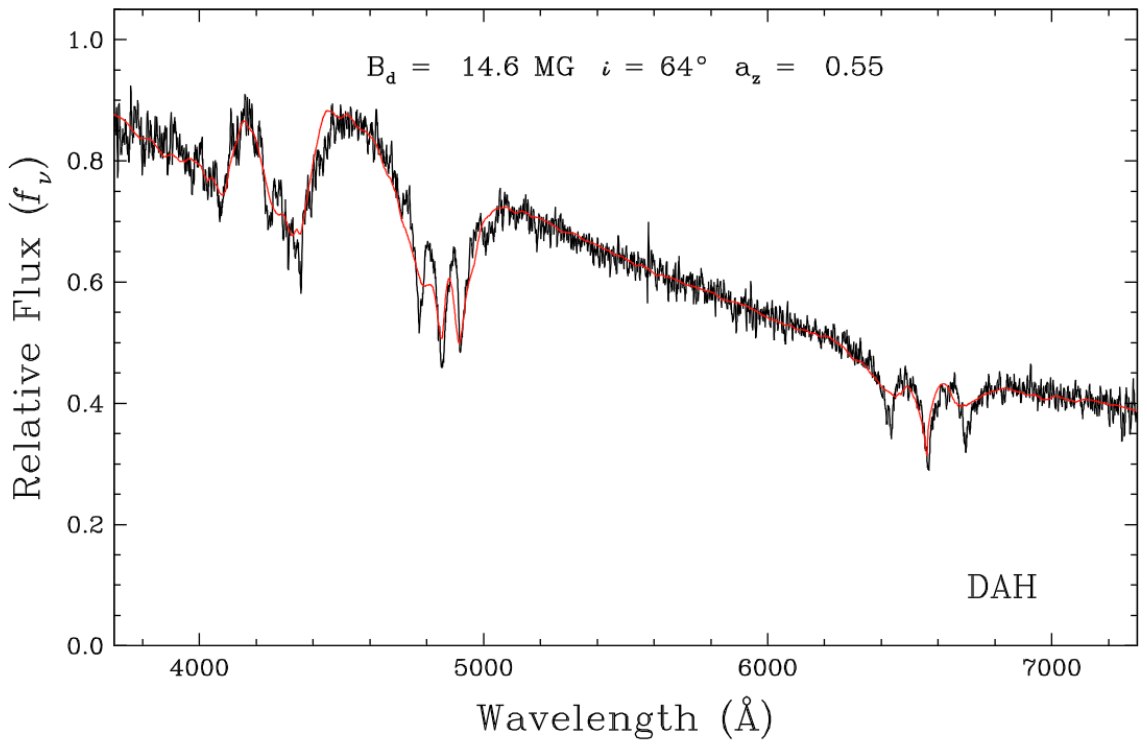}
    \includegraphics[width=3.5in, clip=true, trim=1in 3in 1in 3.4in]{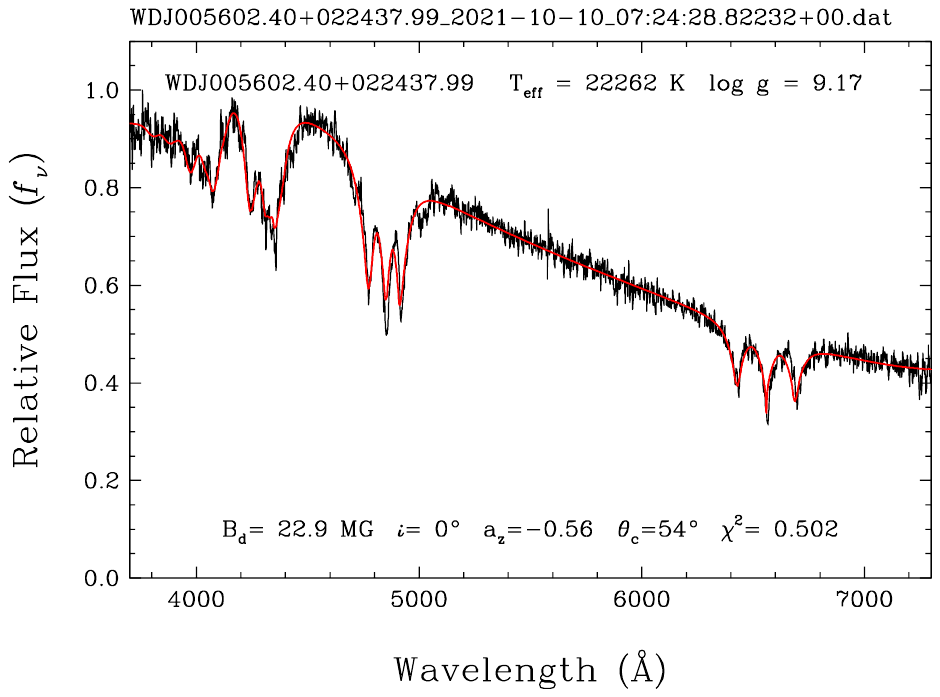}
    \caption{Fit to a hot DAH spectrum with sharp absorption features using the homogeneous model (top panel) and our patchy atmosphere model (bottom panel). The patchy fit is significantly improved despite the unrealistic geometry of a featureless belt.}
    \label{fig4} 
\end{figure}

Another important caveat of this patchy model is that it assumes a H cap and a He region, the latter of which does not contribute to the line opacities. However, not every MWD that requires a patchy fit is below $T_{\rm eff}= 11,000$ K. Figure \ref{fig4} shows the spectrum of a hot DAH, WDJ005602.40+022437.99, with sharp absorption features that our homogeneous model fails to fit, and the improved fit using our patchy atmosphere model. Here the patchy model still assumes an H polar cap and a He belt. However, it now acts as a proxy to restrict the regions of the surface that contribute to the line opacities and thus the range of field strengths sampled by the spectrum. The homogeneous model fails to reproduce the narrow Zeeman components, but interestingly it does at least reproduce the approximate overall strength of the Balmer lines. Hence the patchy solution need not be interpreted as evidence for actual chemical inhomogeneities, rather it simply acts as a filter for which regions of the surface contribute to the spectrum. 

\begin{figure*}
    \centering
    \includegraphics[width=7in, clip=true, trim=0in 0in 0in 0in]{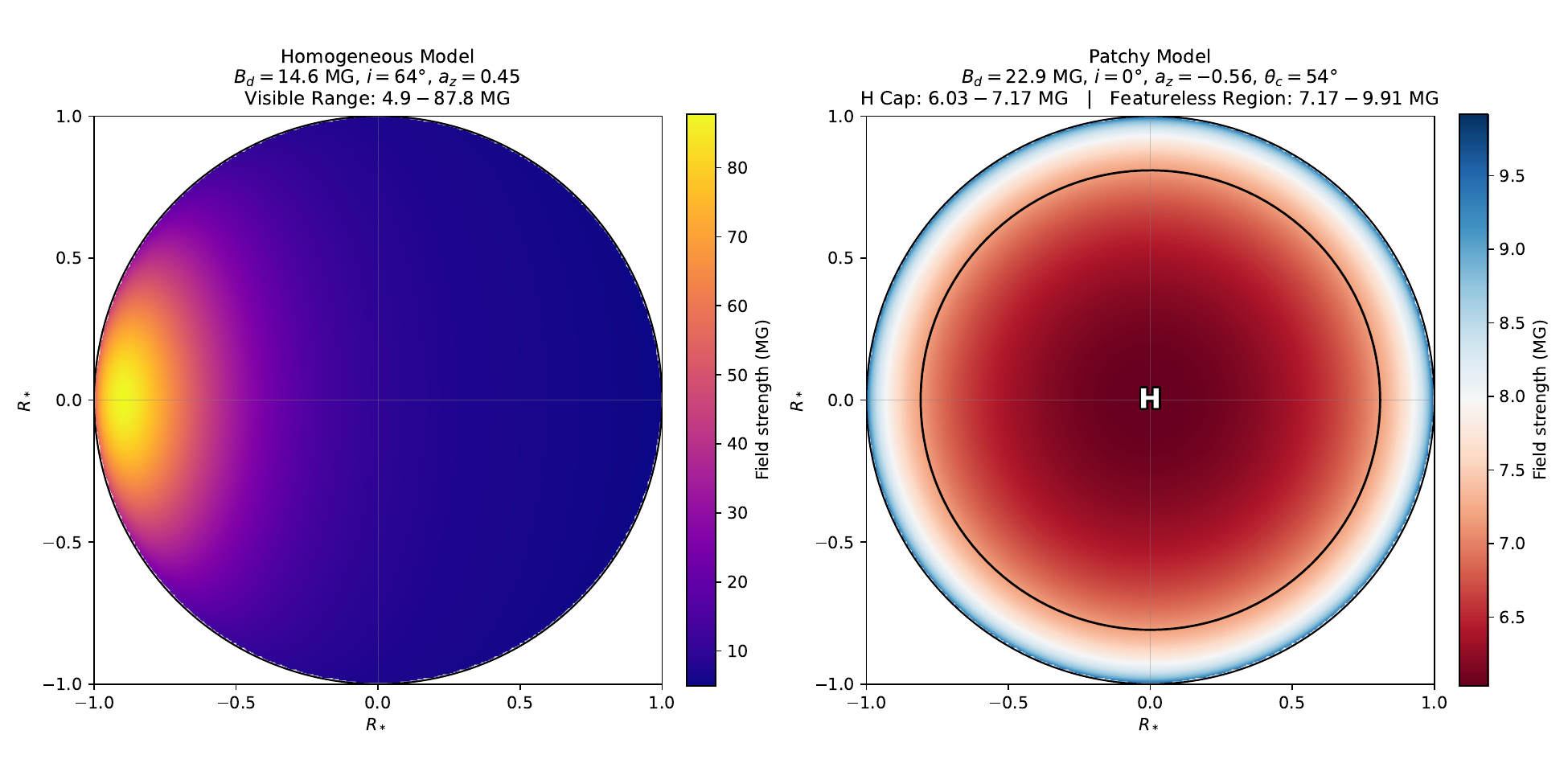}
    \caption{Topology map of WDJ005602.40+022437.99 using our best-fit homogeneous H solution (left panel) and patchy atmosphere solution (right panel). The homogeneous solution produces a nearly edge-on view of the pole, with a wide range of field strengths across the surface. The patchy solution produces a pole-on view and a significantly offset dipole away from the observer. This results in the polar cap having a weak local field strength compared to the remainder of the visible stellar disk. The H polar cap contributes to the line opacities in our fit while the remainder does not.}
    \label{figj0056map}
\end{figure*}

We show the topology map of WDJ005602.40+022437.99 in Figure \ref{figj0056map}. In the patchy case, the observer looks directly at the polar cap, however the dipole offset is $a_z=-0.56$, away from the observer along the z-axis. This large asymmetry results in a lower field strength range across the visible disk than the best fit dipole field strength of 22.9 MG, in contrast to WDJ044543.74$-$031346.97. 

In principle, our patchy atmosphere model should return the same fit as our homogeneous model when $\theta_c=90\degree$. To confirm this, we selected a DAH with a spectrum that is well-fit by the homogeneous model, and compared the solution to that of the patchy model with a forced cap size of $\theta_c=90\degree$. Indeed we obtain the exact same model spectrum between the homogeneous and patchy model. If we allow all four parameters to vary in order to determine the best dipole values and cap size, we obtain similar values using the patchy model as the homogeneous model: $B_d=36$ MG, $i=60\degree$, and $a_z=0.70$ with a large cap size of $\theta_c=78\degree$. 

\subsubsection{Hot vs. Cold Objects}

Among our hot objects, the homogeneous model generally matches the overall strength of the Balmer features. These features are quite strong as expected for white dwarfs at these effective temperatures. This is in contrast to the cool objects, in which the observed features tend to be much weaker than what is expected. Hence the homogeneous model already matches the global Balmer absorption, but it cannot replicate the magnetic field morphology among these hot objects that we need the patchy model for. If we include He I lines in the patchy calculation, not only are He lines generated that do not appear in the spectra, but the fitting procedure is forced to compromise between suppressing those unwanted He features and reproducing the observed H lines, and the resulting solution produces H lines that are too weak. 

Thus there could be two distinct cases of this patchy model based on $T_{\rm eff}$. For hotter objects, since the homogeneous model reproduces the line strength within reason, the model only restricts the region of the surface in order to better reproduce the field structure. For colder objects, since the line strength is generally not well approximated by the homogeneous fits, a He region is necessary to dilute the H line strength. Restricting the relevant region to only a H cap fixes both the Balmer strength and the field geometry issue. In turn, the physical origin of these unique atmospheres could also be distinct. Many white dwarfs undergo convective mixing \citep{Tremblay08,Chen11,Rolland18,Cunningham20,Bedard22,Bergeron22} between $14,000 \geq T_{\rm eff} \geq 6,000$ K in which the convection zone expands to reach the underlying He envelope, switching the dominant atmospheric composition from H to He. However, it is currently unclear if a magnetic field can inhibit this mixing process and generate inhomogeneities on the surface \citep{Cunningham21,Ginzburg25}. 

We also identify two DBAH which we cannot fit with homogeneous atmosphere models \citep[see their Figures 15 and 16]{Kilic26hot}. We obtain excellent fits to these objects using our patchy atmosphere model. \citet{Munday26} discovered three non-magnetic ``double-faced'' white dwarfs in the DESI sample by using a neural network to assign spectral types and then searched for changing assignments between individual spectra. Time-resolved spectroscopy on all five objects would be valuable in order to fully characterize the geometry and nature of each system, as only a few have been studied in detail \citep{Moss25a}. 

Table \ref{tab2} summarizes the parameters of all the MWDs in our sample. We provide the best-fit values from our magnetic modeling procedure, and we specify if we use the homogeneous model or the patchy model for a given object. All of our spectral fits are available to download on Zenodo at \url{https://doi.org/10.5281/zenodo.24112480}

\begin{deluxetable*}{cccccccccc}
\tabletypesize{\footnotesize}
\tablecolumns{10} \tablewidth{0pt}
\tablecaption{Physical parameters for the MWDs in DESI DR1. The full table is available in the online version of this article.} \label{tab2}
\tablehead{\colhead{WD name} & \colhead{Gaia DR3 ID} & \colhead{Spectral Type} & \colhead{$T_{\rm eff}$} & \colhead{Mass} & \colhead{Cooling Age} & \colhead{$B_d$} & \colhead{$i$} & \colhead{$a_z$} & \colhead{Model Fit} \\
& & & (K) & ($M_\odot$) & (Gyr) & (MG) & (\degree) & ($R/R_{\odot}$) & }
\startdata
J0002$+$0733 & 2745919106553695616 & DAH & $8011\pm70$ & $0.786\pm0.015$ & $1.85\pm0.10$ & 2.6 & 83 & -0.17 & Patchy\\
J0002$-$1209 & 2421562244351016192 & DAH? & $11456\pm401$ & $0.724\pm0.108$ & $0.585\pm0.122$ & 126.2 & 2 & 0.25 & Homogeneous\\
J0006$+$0755 & 2746037712074342784 & DAH & $8712\pm71$ & $0.746\pm0.018$ & $1.28\pm0.05$ & 11.4 & 25 & 0.62 & Homogeneous\\
\enddata

\end{deluxetable*}

\subsubsection{Previous Fits from the Literature}

\begin{figure}[!ht]
    \centering
    \includegraphics[width=3.5in, clip=true, trim=0in 2.75in 0in 2in]{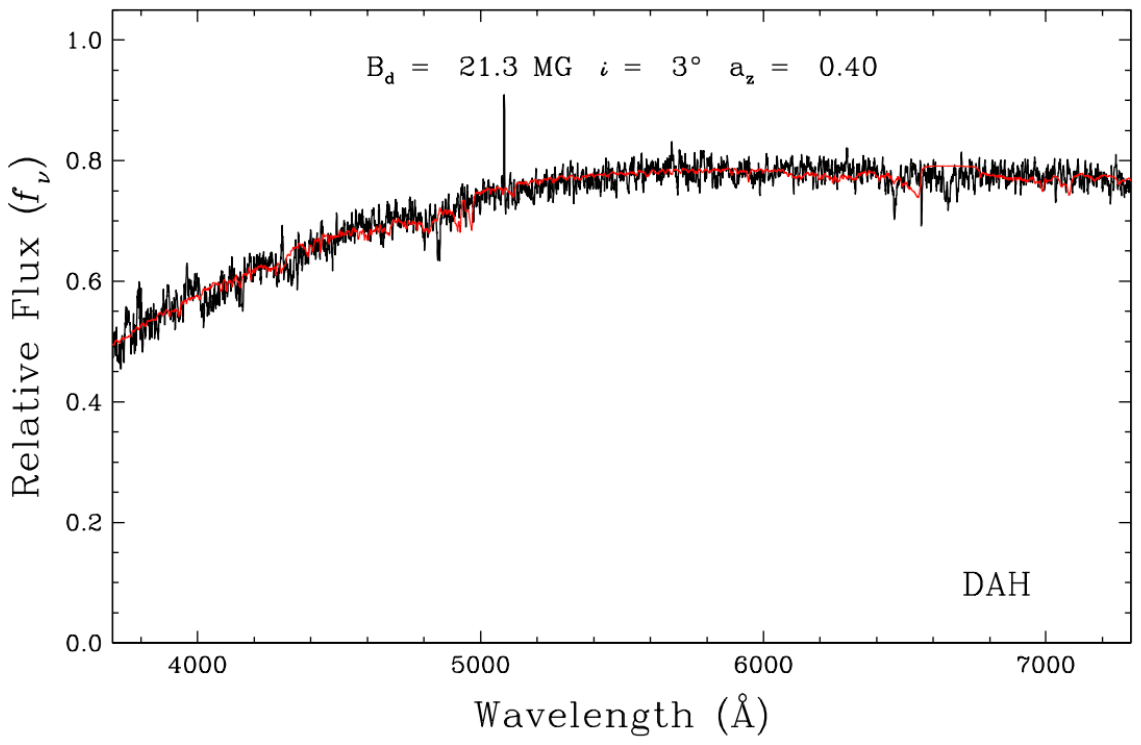}
    \includegraphics[width=3.5in, clip=true, trim=0.85in 1in 0.55in 1in]{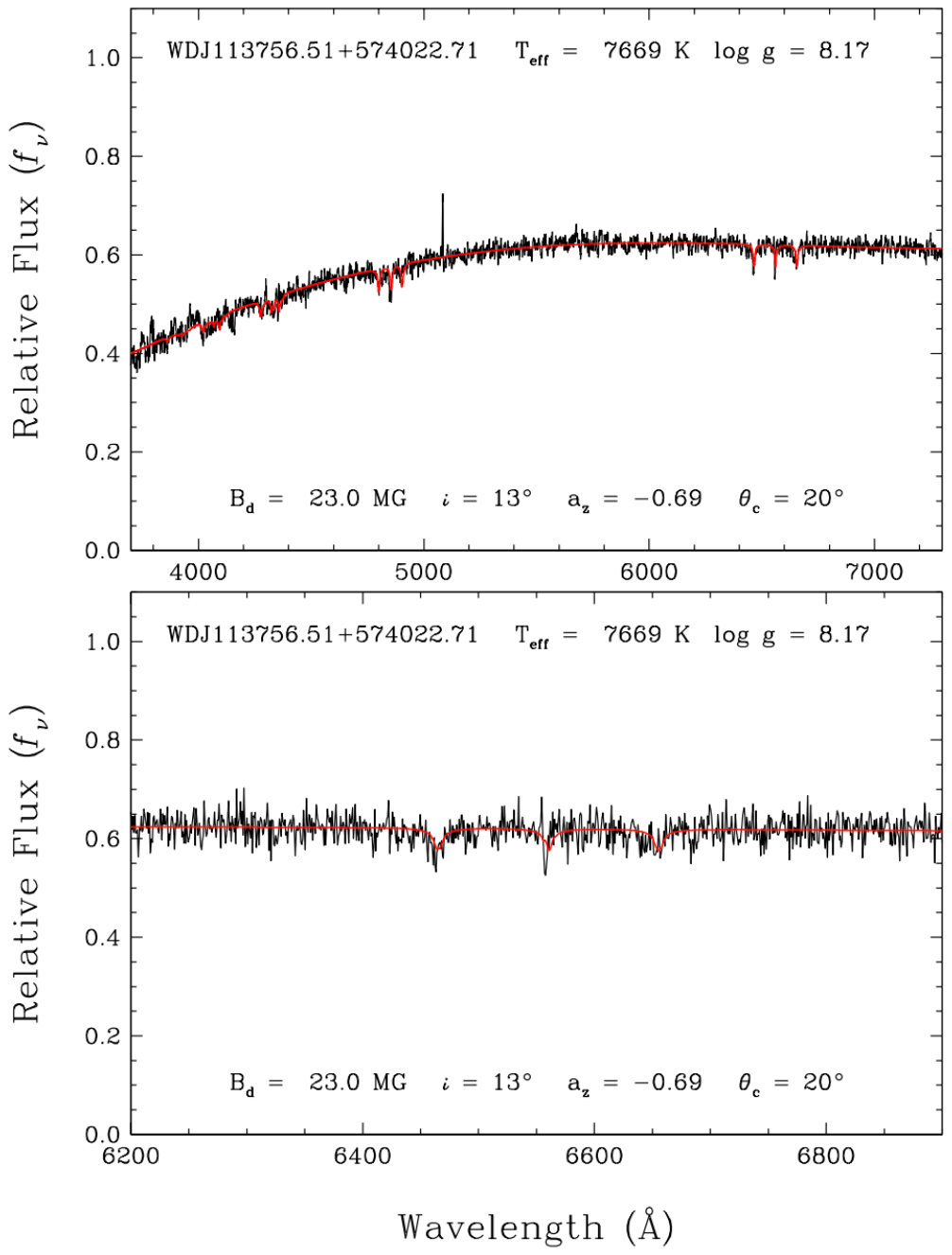}
    \caption{Top: Homogeneous fit to a DAH, WDJ113756.51+574022.71. The fit fails to reproduce the shallow Balmer lines. Middle: Patchy fit to the entire spectrum using the parameters obtained from the fit to the H$\alpha$ region. Bottom: Patchy fit to the H$\alpha$ region.}
    \label{fig13} 
\end{figure}

Issues with fitting absorption features have previously been reported in the literature similar to the ones presented here. \citet{Kulebi09} analyzed 141 DAHs from SDSS and identified several objects with sharper or shallower than expected line profiles, indicating more complex geometries in those objects. \citet{Hardy23a} obtained fits for 251 DAHs and could not obtain suitable fits for 111 of them, while \citet{Hardy23b} analyzed 43 DBHs but could only obtain suitable fits for eight of them. Several of the MWDs that require patchy fits in our analysis have been previously fit with homogeneous models. The spectrum of WDJ113756.51+574022.71 possesses shallow features, such that neither a centered nor offset dipole model successfully fit the spectrum \citep[see their Figure 1]{Kulebi09}. Figure \ref{fig13} shows our homogeneous and patchy fits to this object. Similar to \citet{Kulebi09}, our homogeneous fit fails to reproduce the shallow absorption features. Instead, the patchy model yields an excellent fit to all of the Balmer lines, not just H$\alpha$. We obtain a field strength of 23.0 MG, an order of magnitude higher than the 2.83 MG result from \citet{Kulebi09}. It is worth mentioning that the higher resolution of DESI over SDSS allows for more robust detection and fitting of split features in our analysis compared to previous results, though in the case of this specific object the split lines can be detected in the SDSS spectrum. 

\begin{figure}[!ht]
    \centering
    \includegraphics[width=3.5in, clip=true, trim=0in 2.75in 0in 2in]{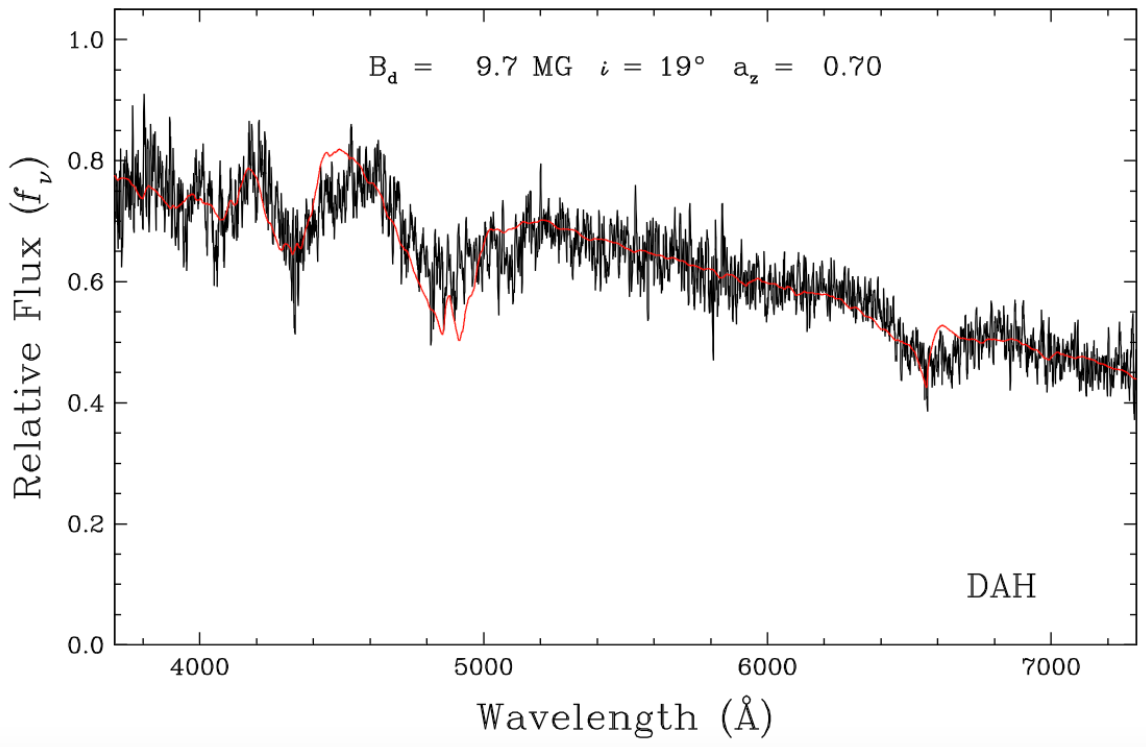}
    \includegraphics[width=3.5in, clip=true, trim=0.85in 3in 1in 3.4in]{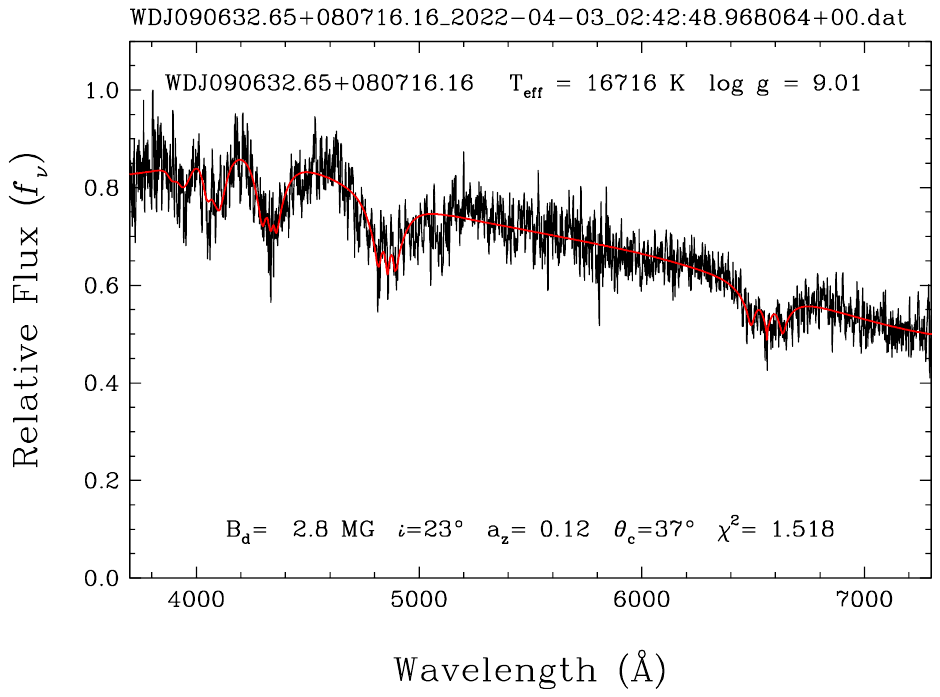}
    \caption{Top: Homogeneous fit to the DAH WDJ090632.65+080716.16.  Bottom: Patchy fit to the entire spectrum.}
    \label{fig14} 
\end{figure}

Another example is WDJ090632.65+080716.16. \citet{Kulebi09} obtained a field strength of 5.97 MG for this object, but encountered similar issues fitting the shallow absorption features. Figure \ref{fig14} shows our homogeneous and patchy fits to this object. Once again, the patchy model provides an excellent fit to the spectrum. Note that this object is at $T_{\rm eff} = 16716$ K, while WDJ113756.51+574022.71 is at $T_{\rm eff} = 7669$ K. Hence if there is a shared mechanism that produces these unique features, it must apply across a broad temperature range. 

\citet{Hardy23a} identified 28 DAHs that have shallow absorption features and could not be fit by a homogeneous offset dipole model (see their Table 3). Of those, 11 of them are in the DESI DR1 sample, 6 of which have improved fits when using a patchy model: WDJ094416.63-001855.53, WDJ113756.51+574022.71, WDJ131426.39+173228.40, WDJ141813.22+312340.03, WDJ151625.10+280320.94, and WDJ222348.40+231909.37. However, it is worth noting that not every object that displays shallow absorption profiles requires a patchy atmosphere in our fits. For example, \citet{Kulebi09} could not fit the shallow lines of WDJ112328.50+095619.35, whereas our homogeneous fit matches the observed features. \citet{Hardy23a} found that WDJ133250.72+011706.32, WDJ163917.47+103604.84, WDJ211504.83+040009.95, WDJ213148.67+065929.57, and WDJ213819.85+112311.34 have shallow features and could not be fit with a homogeneous model, but the homogeneous model achieves satisfactory results in our analysis.  

\begin{figure}[!ht]
    \centering
    \includegraphics[width=3.5in, clip=true, trim=0in 3.5in 0in 2.95in]{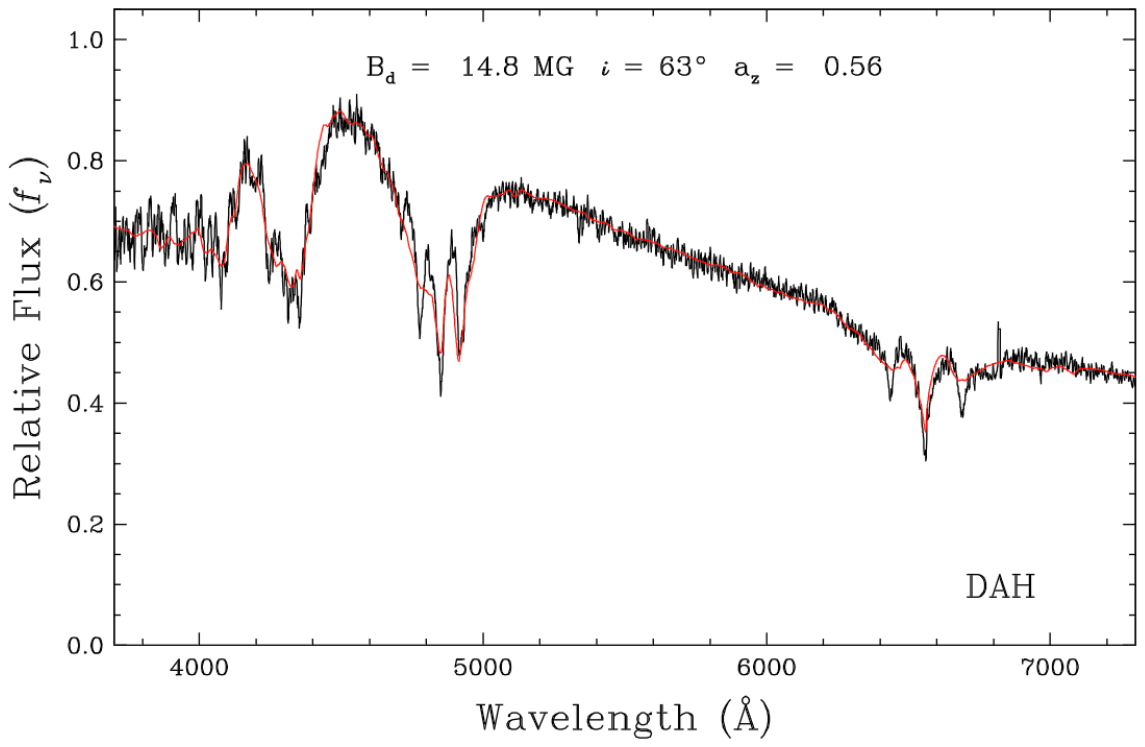}
    \includegraphics[width=3.5in, clip=true, trim=0.3in 2.75in 0.15in 2.75in]{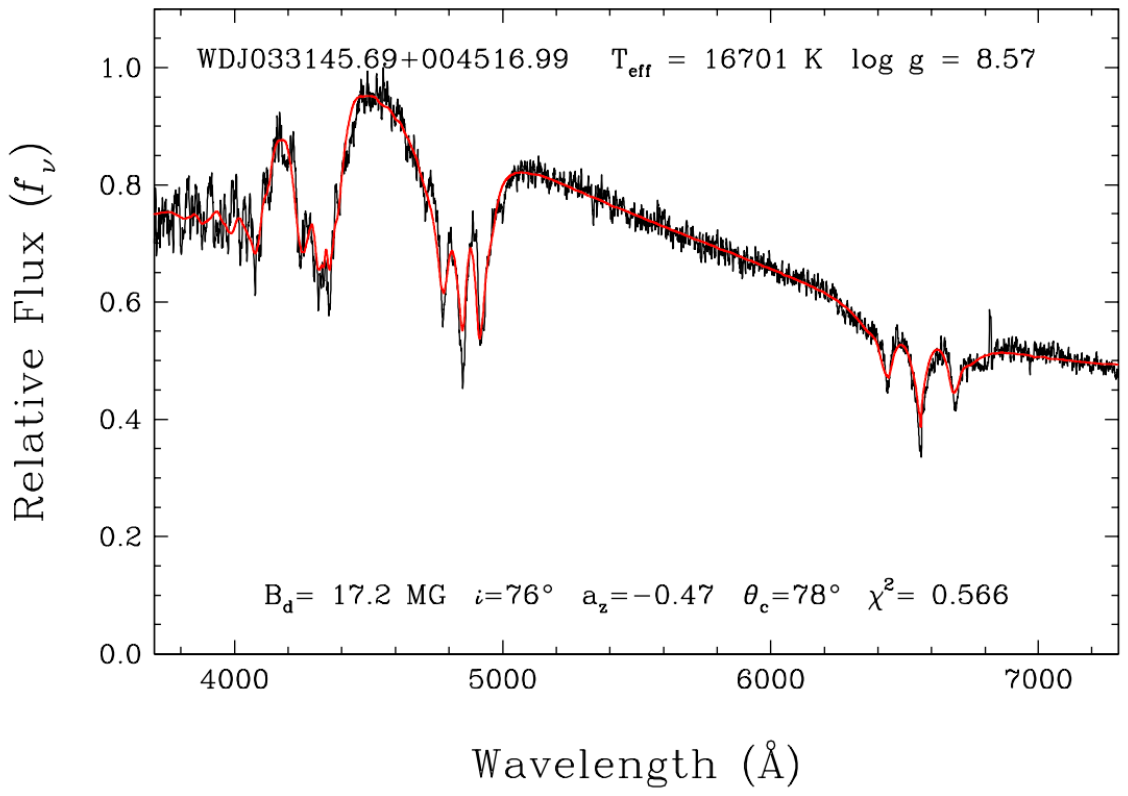}
    \caption{Top: Homogeneous fit to the DAH WDJ033145.69+004516.99.  Bottom: Patchy fit to the entire spectrum.}
    \label{figsharp} 
\end{figure}

In contrast to shallow features, many objects that require patchy fits have sharp line profiles. Figure \ref{figsharp} shows our homogeneous and patchy fit to one of these objects, WDJ033145.69+004516.99. The homogeneous model predicts shallow $\sigma$ components at H$\alpha$ and H$\beta$, but the observed features are instead sharp and pronounced. The patchy model obtains an excellent fit to this spectrum, with a large cap size of $\theta_c=78\degree$. \citet{Hardy23a} previously fit this object and classified it as a ``Bad Dipole Fit'' (see their Table 6). However, \citet{Kulebi09} achieved a good fit using a homogeneous model. 

Lastly, many of these patchy objects are likely rapid rotators, such that the 15$-$20 minute long exposures from DESI could sample multiple regions of the complex field distribution. This would produce unique features such as 5 Zeeman components, which can be seen in the H$\beta$ feature of WDJ033145.69+004516.99 in Figure \ref{figsharp}. An extreme example is WDJ113357.66+515204.69, which \citet{Manser24} found to have five distinct profiles at H$\alpha$ in its DESI spectrum. They modeled the DESI spectrum as a superposition of two sets of Zeeman profiles and determined two field strength estimates of 3.04 and 4.4 MG. To determine if this target is in fact a rapid rotator, we obtained 18 spectra on this target at the 6.5m MMT with the blue channel spectrograph. We obtained two series of back-to-back exposures, one with 300 s long exposures and the other 240 s long exposures; the first series was obtained on UT 2024$-$05$-$01 while the second was taken on UT 2024$-$05$-$02. We used the 1.25'' slit with the 500 l mm$^{-1}$ grating which yields a resolution of 4.5 \AA. 

\begin{figure*}[]
    \centering
    \includegraphics[width=3.3in, clip=true, trim=0.35in 0.75in 1in 1.4in]{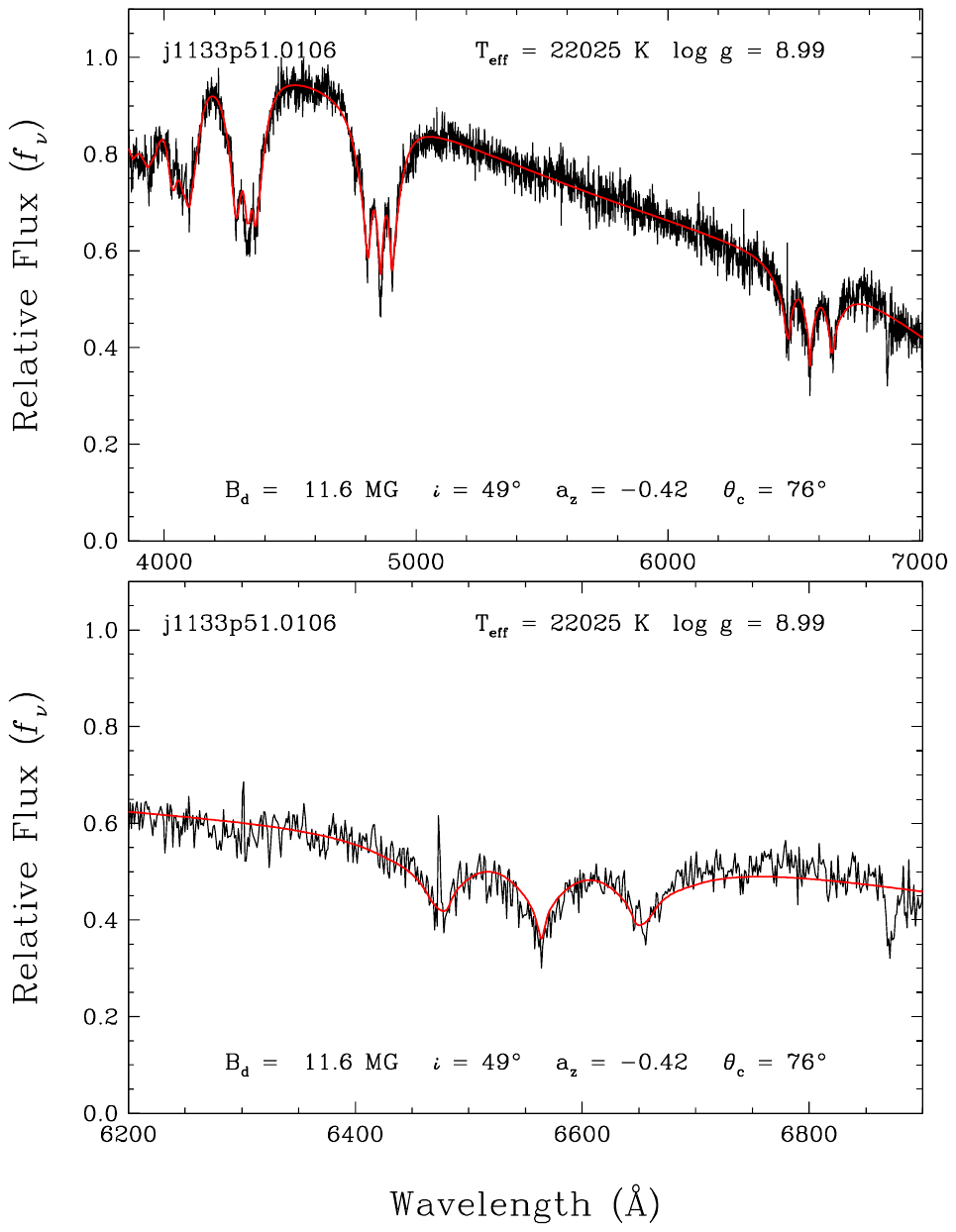}
    \includegraphics[width=3.3in, clip=true, trim=0.35in 0.75in 1in 1.4in]{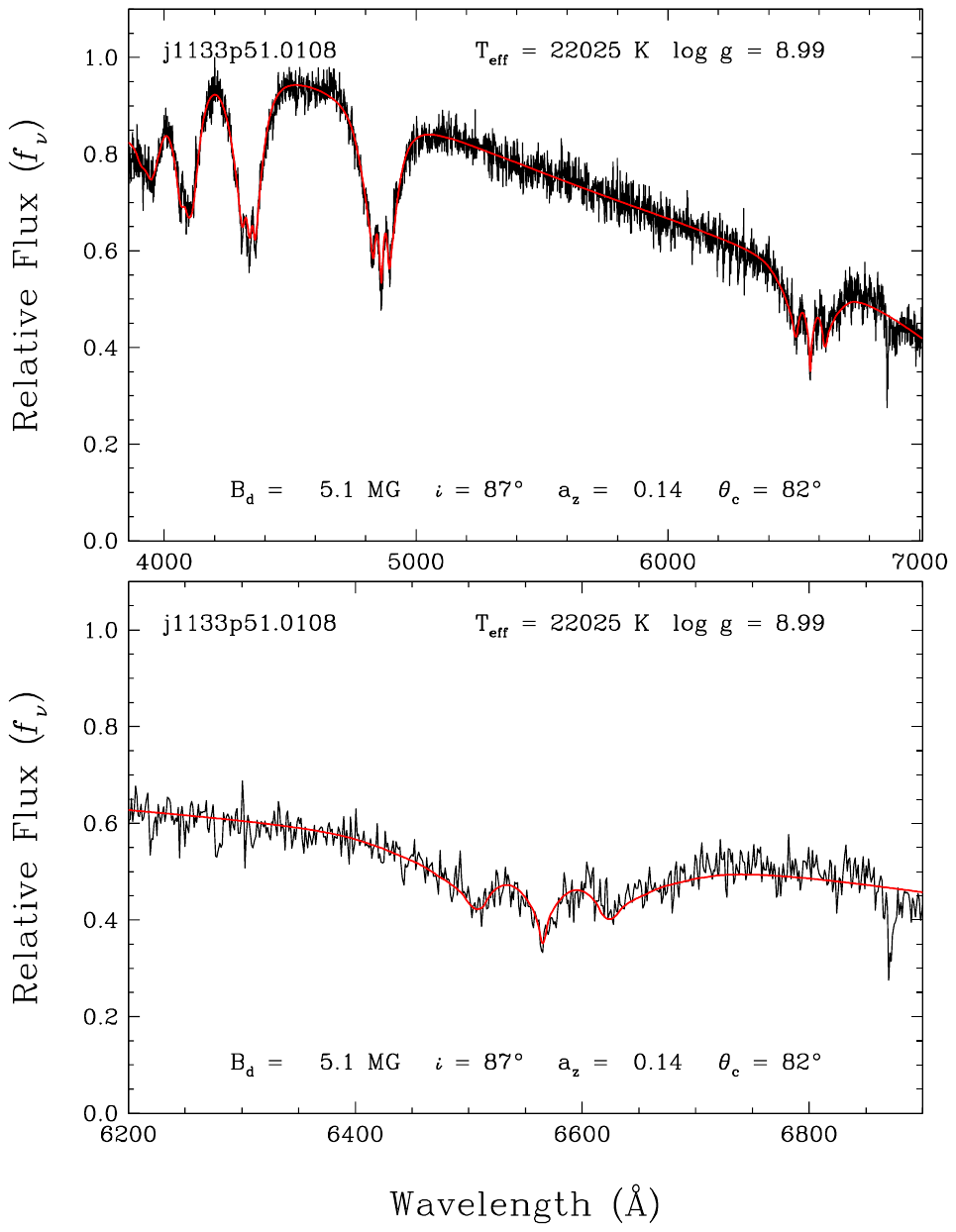}
    \caption{Patchy fits to two MMT spectra of a DAH that shows five Zeeman components in its DESI spectrum, WDJ113357.66+515204.69. The line profiles change significantly between the two spectra, with more pronounced profiles in the left spectrum that are further from the central component compared to the spectrum on the right. The differences in the spectral features result in different magnetic parameters from our fits.}
    \label{figj1133}
\end{figure*}

Figure \ref{figj1133} shows our patchy fits to two of the MMT spectra separated by 10 minutes. The MMT spectra show notable changes in the line profiles, and the patchy model yields excellent fits to both spectra. We obtain significantly different field strengths of 11.6 and 3.5 MG due to the observed changes in the spectral features. Hence this target has a rotation period comparable to or shorter than that of the DESI exposure length. MWDs with unusual line profiles have turned out to be rapid rotators \citep{Kilic19,Moss23} with complex field structures and potentially inhomogeneous atmospheres. Further time-series analysis on WDJ113357.66+515204.69 and other variable MWDs from DESI will constrain their rotation periods and provide valuable insight into the structure of their magnetic fields. 

\subsection{Asymmetric Line Profiles}

Two objects with unusual asymmetric Balmer-line profiles,
WDJ010338.56$-$052251.96 (J0103$-$0522) and
WDJ115952.05+000751.87 (J1159+0007), have been discussed in
several previous studies, with their interpretation evolving as
additional observations and model comparisons became available.
J1159+0007 was originally discussed by \citet{Limoges15} as a
possible unresolved double-degenerate system because of the
discrepancy between its photometric and spectroscopic temperatures.
\citet{Tremblay20} subsequently investigated J0103$-$0522, whose
Balmer lines show similar asymmetric profiles and wavelength shifts,
and showed that the shifts of the Balmer-line minima are consistent
with the quadratic Zeeman effect. They obtained a characteristic
field strength of 4.82 MG, while cautioning that this value should
not be interpreted as the global mean surface field strength.

A different interpretation was considered by \citet{Caron23}, who
noted that J0103$-$0522 and J1159+0007 are nearly identical
spectroscopic twins. The unusually broad H$\alpha$ profiles could
be reproduced with He-dominated atmospheres containing traces of
hydrogen, in which van der Waals broadening by neutral helium
significantly affects the H$\alpha$ profile. This led to their
interpretation as He-rich DA stars, and \citet{Caron23} suggested
that the unusual line profiles might result from helium broadening
not fully accounted for in the models. More recently,
\citet{Kilic25} reconsidered these solutions by examining the
complete Balmer spectra. Although the He-rich models reproduce
H$\alpha$, they predict the higher Balmer lines to be much too weak.
The He-rich interpretation was therefore rejected in favor of a
pure-H DA solution, although the origin of the unusual line profiles
remained unexplained.

In the present analysis, we have revisited the pure-H solution over
the entire Balmer spectrum. This reveals the complementary problem:
while the He-rich solution predicts the higher Balmer lines to be
much too weak, the pure-H solution predicts them to be much too
strong. Thus, neither of the homogeneous atmospheric compositions
previously considered can simultaneously reproduce H$\alpha$ and
the higher members of the Balmer series. This discrepancy motivated
us to explore whether an inhomogeneous surface composition could
reconcile the different regions of the spectrum.

\begin{figure*}[]
    \centering
    \includegraphics[width=2.3in, clip=true, trim=1in 2.35in 1in 3.35in]{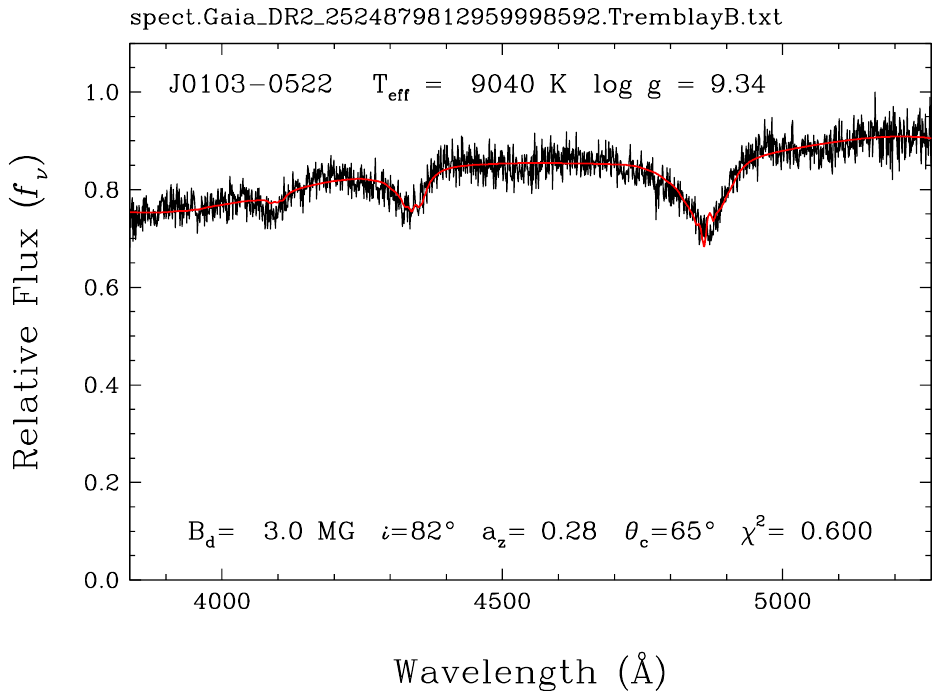}
    \includegraphics[width=2.3in, clip=true, trim=0.3in 2.3in 0.6in 3.3in]{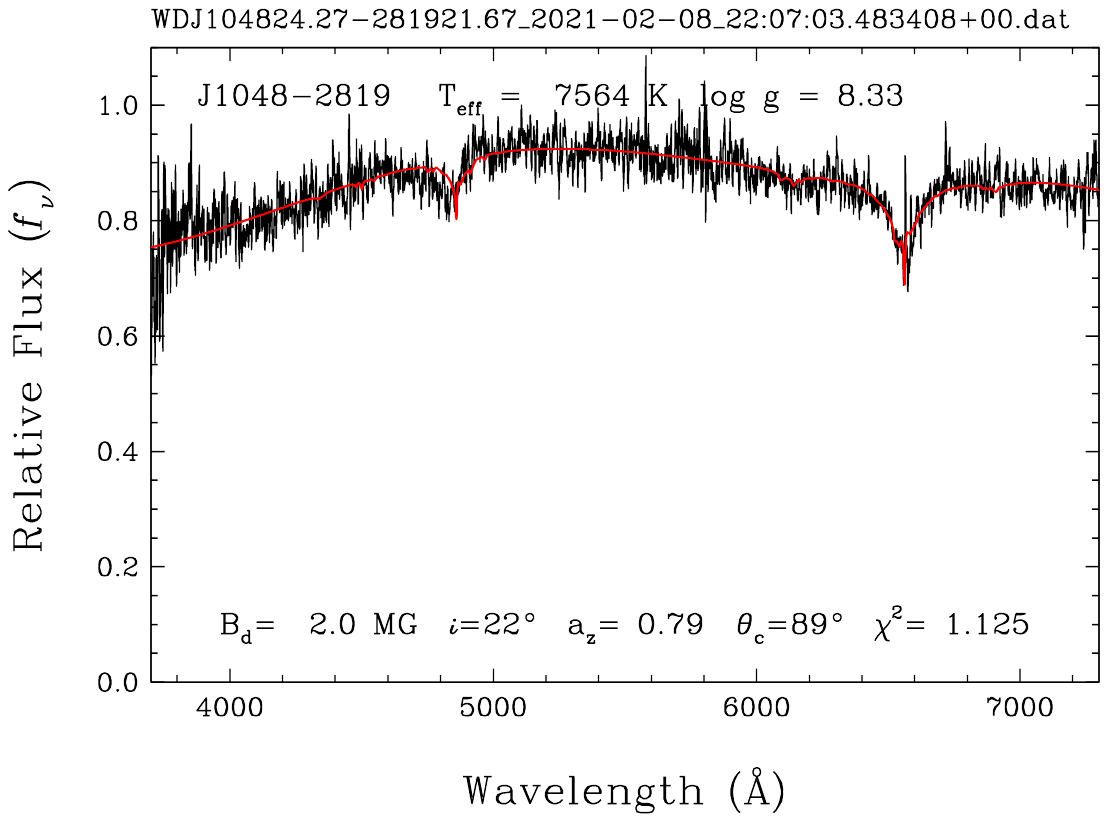}
    \includegraphics[width=2.3in, clip=true, trim=1in 2.5in 1in 3.4in]{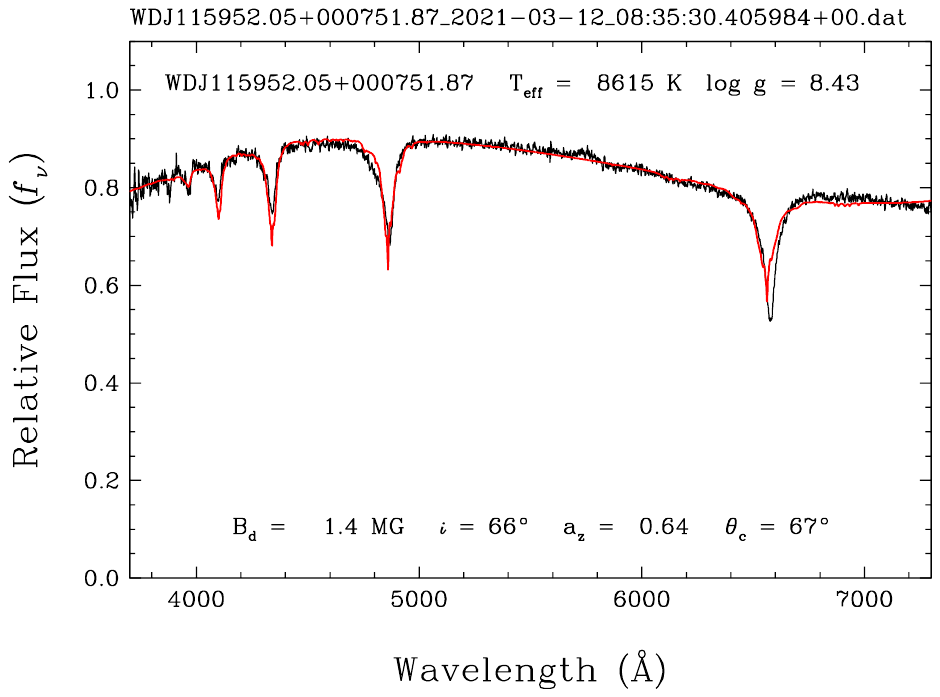}
    \caption{Patchy fits to three MWDs with asymmetric line profiles.
    The spectra for J1048$-$2819 and J1159+0007 are from DESI, while
    the spectrum for J0103$-$0522 is from \citet{Tremblay20}.}
    \label{figasym}
\end{figure*}

We first explored this possibility for J1159+0007 using our patchy
magnetic model. We find that the Balmer spectrum can be reproduced
when the polar caps themselves contain a mixed H/He composition,
while the remainder of the surface is He-rich. Importantly, the
mixed H/He caps were not assumed from the outset, but emerged from
our attempts to reproduce simultaneously H$\alpha$ and the higher
Balmer lines. We then applied the same prescription to
J0103$-$0522 and also obtained an excellent fit. Figure
\ref{figasym} shows the resulting fits to these two objects.

We also identify a third object in the DESI sample,
WDJ104824.27$-$281921.67 (J1048$-$2819), with an unusual asymmetric
line profile. Figure \ref{figasym} also shows our fit to this object,
which remains problematic. J1048$-$2819 was initially classified as
a He-DA, but its H$\beta$ feature is shifted relative to the expected
position. A patchy model with H-rich caps fails to reproduce this
feature and converges to $\theta_c=89\degree$, effectively
eliminating the patch. A homogeneous offset-dipole model with a
mixed H/He atmosphere similarly fails to reproduce the H$\beta$
profile. The origin of the peculiar line profile in J1048$-$2819
therefore remains uncertain; an inaccurate effective temperature
may contribute to the discrepancy.

Our results provide a magnetic, compositionally inhomogeneous
interpretation of J0103$-$0522 and J1159+0007 that can reproduce
their unusual Balmer spectra without invoking an unresolved binary
system or an additional helium-broadening mechanism. We emphasize,
however, that the success of the mixed H/He patchy model does not
uniquely establish the actual surface abundance distribution.
Rather, it demonstrates that a weak magnetic field combined with an
inhomogeneous line-forming surface can simultaneously account for
H$\alpha$ and the higher Balmer lines in these objects.

\subsection{DAHe White Dwarfs}

\begin{figure}[!ht]
    \centering
    \includegraphics[width=3.5in, clip=true, trim=0in 3.4in 0in 3in]{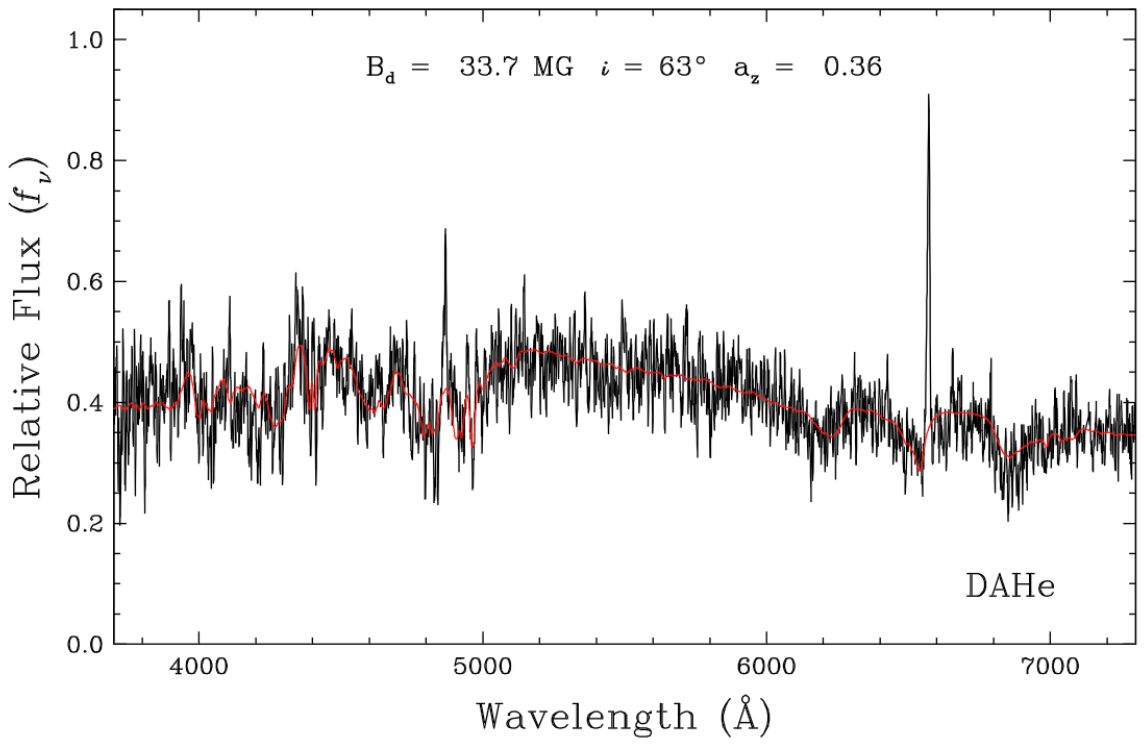}
    \includegraphics[width=3.5in, clip=true, trim=0in 3in 0in 3in]{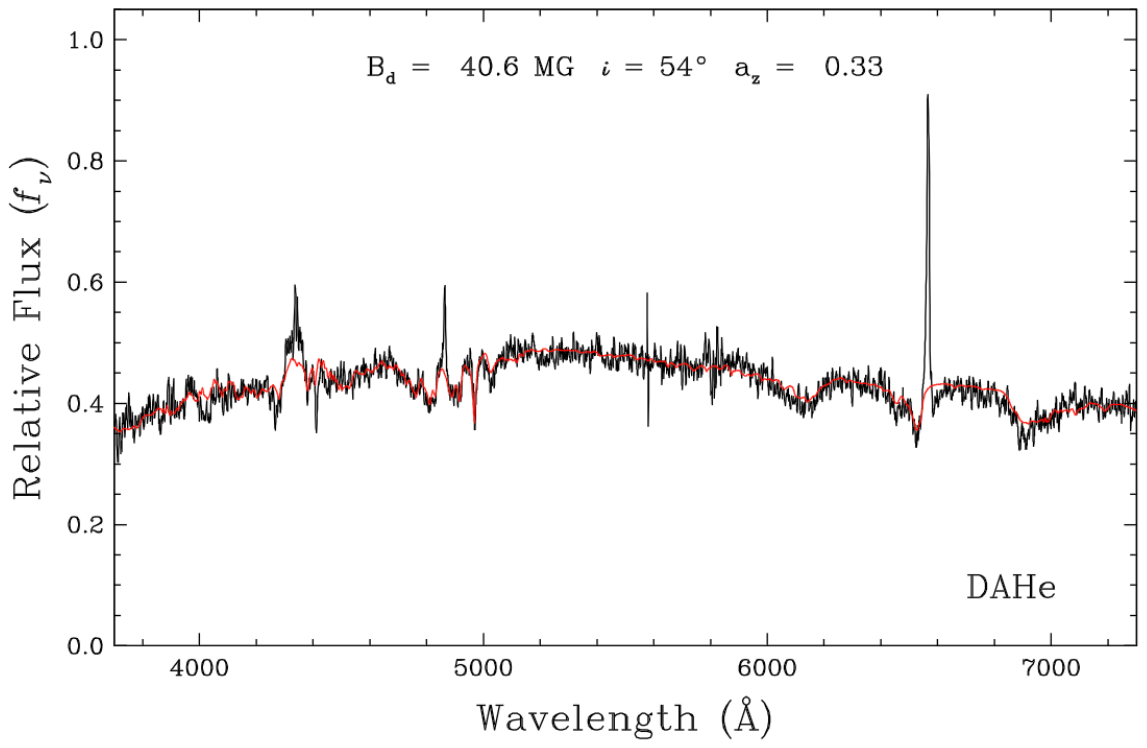}
    \caption{Dipole fits to two of the DAHe white dwarfs in the DESI DR1 sample, WDJ111257.59+690255.98 (top panel) and WDJ151415.68+074446.75 (bottom panel). We achieve excellent fits despite the fact that the model does not include emission features.}
    \label{figdahefits} 
\end{figure}

We identify 37 objects with Zeeman-split Balmer lines in emission \citep{Manser23}. These unusual objects likely exhibit chromospheric activity given the observed antiphase variability between broadband photometry and the strength of the Balmer line emission \citep{Greenstein85,Gansicke20,Reding20,Walters21,Elms23,Manser23,Reding23,Elms26}. We achieve excellent dipole fits for two of these, WDJ111257.59+690255.98 and WDJ151415.68+074446.75. Figure \ref{figdahefits} shows our dipole fits to these two objects. While our model does not include emission features, it does an excellent job fitting the split Balmer lines. These two objects have relatively high field strengths at 33.7 and 40.6 MG respectively. \citet{Manser23} identified 19 DAHe white dwarfs and 14 candidates in DESI DR1. All but one of these objects, WDJ004736.08$–$154326.57 which is a candidate, are classified as DAHe white dwarfs in our analysis. While these objects occupy a narrow range of cooling ages, the field strengths determined by \citet{Manser23} range from 5 MG to 147 MG, indicating a potentially complex origin for these objects.

\subsection{Problematic Objects}

Here we briefly mention 12 objects that we cannot reproduce with our homogeneous or patchy model. These are J0227$-$0553, J0333+0007, J0648+5213, J0656+6550, J0720+7127, J0842$-$0222, J1046$-$0518, J1202$-$0324, J1303$-$0852, J1338$-$0432, J2138+1123, and J2200$-$0614. Several of these have strong fields given the distortion of their spectral features, such that fitting the spectra is unsurprisingly difficult. The spectrum for J0227$-$0553 shows numerous spikes that are likely not real, which makes fitting the spectrum challenging. Three objects show a large feature near 5900 \AA\ that is not included in our model: J0656+6550, J1202$-$0324, and J2200$-$0614. Finally, J2138+1123 displays what appear to be normal Balmer lines in addition to shifted features that are expected from a strong field. Our homogeneous model can fit the shifted features but does not reproduce the Balmer features. This could be a DA+DAH binary system, however we label it as a DAH: for now. 

\section{Results}

\subsection{Sample Findings}

\begin{figure*}[!ht]
    \centering
    \includegraphics[width=7in, clip=true, trim=0in 0in 0in 0in]{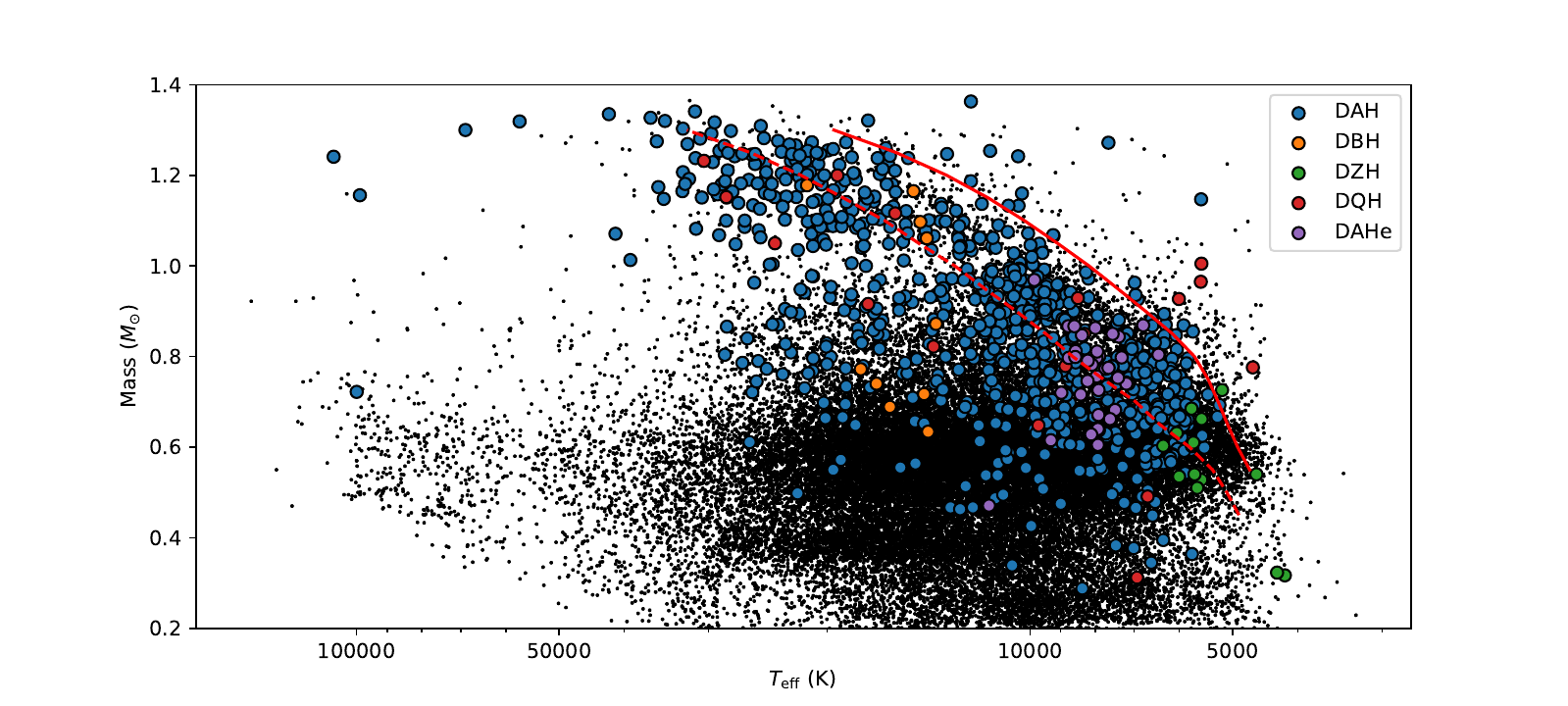}
    \caption{Photometric masses as a function of effective temperature. Black points are non-magnetic objects, while the MWDs are colored by spectral type. The dashed red line corresponds to the onset of crystallization while the solid red line corresponds to 80\% core-crystallization. DAHs (including DAH: and DAH?) dominate the sample, and many systems, particularly between 0.4 and 1.0 $M_{\odot}$, have not begun crystallizing. DAHes are found almost exclusively between 6,000 and 10,000 K, while DZHs are only found below 7,000 K.}
    \label{fig7} 
\end{figure*}

Despite the biases against faint objects and low field strengths, the large number of objects from DESI DR1 allows for the study of white dwarf magnetism at an unprecedented level. Figure \ref{fig7} shows the distribution of photometric masses as a function of effective temperature for the entire DESI WD sample, with MWDs colored by spectral type. It is clear the source of magnetic fields in many objects cannot be explained by a crystallization-induced dynamo \citep{Isern17,Schreiber21,Ginzburg22}, as they are too hot to have begun core crystallization. 

\begin{figure}[!ht]
    \centering
    \includegraphics[width=3.5in, clip=true, trim=0in 0in 0in 0in]{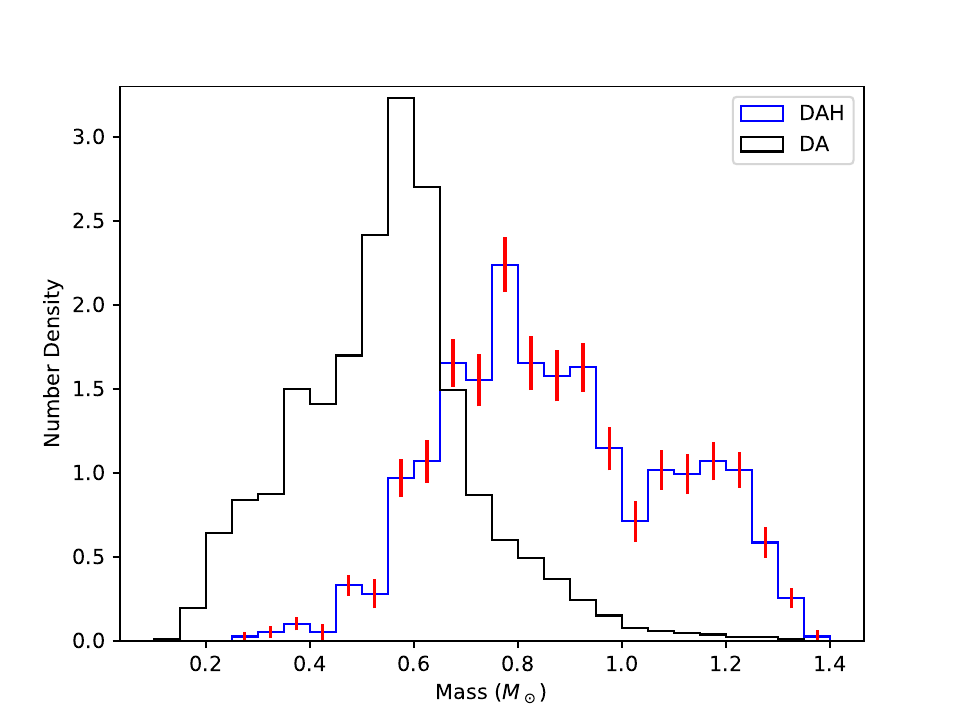}
    \caption{Mass distributions of magnetic and non-magnetic DAs in the DESI sample. There is a steep drop-off in DAs after 0.6 $M_{\odot}$, while the majority of DAHs occur above 0.6 $M_{\odot}$.}
    \label{fig8} 
\end{figure}

Figure \ref{fig8} shows the mass distribution of the magnetic and non-magnetic white dwarfs with hydrogen-dominated atmospheres. To calculate the error in each mass bin for the MWDs, we draw new masses from a Gaussian distribution centered on each mass value with a standard deviation equal to the mass error, rebuild the histogram, and then repeat this process 10,000 times. The association between magnetism and higher mass WDs is well-known \citep{Liebert03,Kepler13,Ferrario15,Mccleery20,Amorim23,Obrien24,Moss25b}. This can partially be explained by binary evolution products that produce hot, massive systems \citep{Obrien24,Jewett24,Moss25b}. However, the median of the DAH mass distribution is 0.87 $M_{\odot}$, below the canonical ultramassive regime ($M\geq0.9$ $M_{\odot}$). Additionally, \citet{Temmink20} performed a population synthesis study on the single WDs within 100 pc and found that 30$-$50\% of ultramassive systems come from binary mergers, while 10$-$30\% below this mass range do. Hence while mergers are common especially among high mass white dwarfs, single-star evolution products are preferentially common even in the ultramassive regime. So unless stellar mergers produce the overwhelming majority of MWDs across a wide mass range, which is unlikely, single-star evolution generates MWDs at higher masses than non-magnetic white dwarfs. 

Interestingly, the errors in the mass bins suggest a decrease in the number density of MWDs at $\sim$1.0 $M_{\odot}$. \citet{Bagnulo22} found that only one out of six WDs between 0.95 and 1.10 $M_{\odot}$ are magnetic in their sample. However, this mass range was incomplete in their sample due to only observing objects younger than 0.6 Gyr between 20 and 40 pc. A similar drop was found by \citet{Amorim26b} who identified 547 MWDs in DESI DR1 (see their Figure 11). \citet{Obrien24} did not find a decrease between 1.0 and 1.05 $M_{\odot}$, however they did not find any MWDs between 0.95 and 1.0 $M_{\odot}$ while finding objects on either side of this mass range. Given the dual formation channel of MWDs \citep{Bagnulo22,Moss25b}, binary evolution may infrequently produce WDs at this mass, while the products of single-star evolution may not yield MWDs at this mass and above. \citet{Berro12} simulated the masses of remnants from different types of mergers, including double WD mergers and evolved stars, and found a lower fraction of merger products from double WD mergers at 1 $M_{\odot}$ compared to 0.9 and 1.1 $M_{\odot}$ (see their Figure 2). Double white dwarf mergers become more common among ultramassive systems, with roughly half of all merger remnants in this mass regime coming from the double degenerate scenario \citep{Temmink20}. Using the IFMR from \citet{Cummings18}, a 1 $M_{\odot}$ WD corresponds to a $\sim$5 $M_{\odot}$ star on the main-sequence. \citet{Stello16} measured the dipole oscillation frequencies in 3600 \textit{Kepler} red giants and found that core magnetic fields, which likely produce magnetism in WDs, suppressed the frequencies in 20\% of the giants. However the suppression fraction varies with mass: $\sim$60\% of stars between 1.6 and 2.0 $M_{\odot}$ show suppressed frequencies, but this drops to $\sim$40\% above 2.0 $M_{\odot}$. Higher mass giants are rare in the \textit{Kepler} field, so precise asteroseismology is typically unavailable to detect core fields in these stars that would produce higher mass WDs. Hence the ability to produce high mass MWDs through single-star evolution, while unlikely, remains an open question.

\begin{figure}[!ht]
    \centering
    \includegraphics[width=3.5in, clip=true, trim=0in 0in 0in 0in]{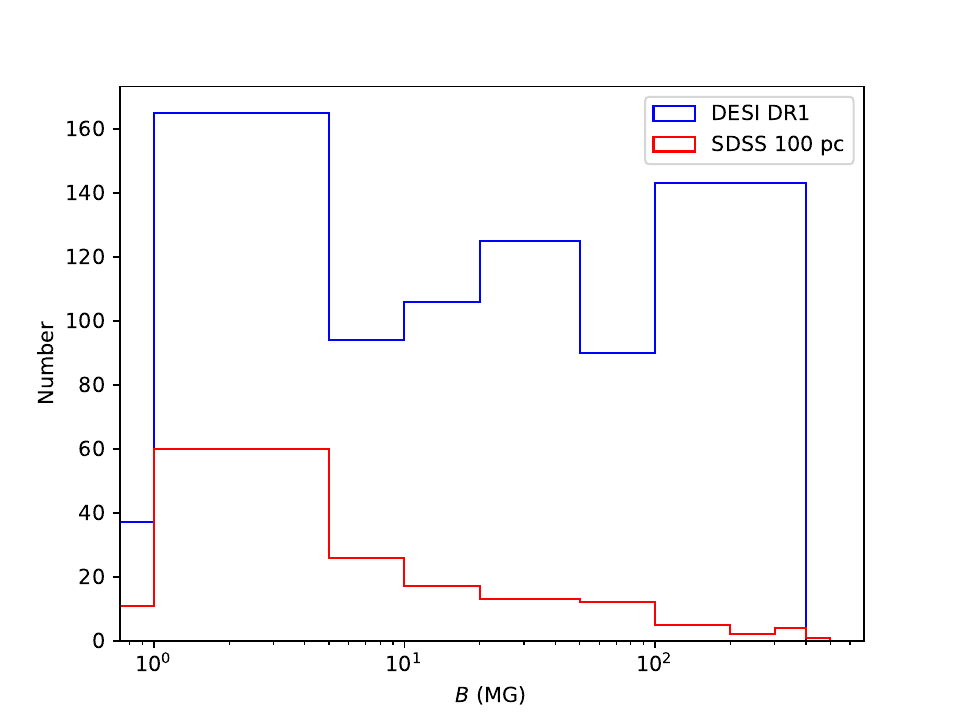}
    \includegraphics[width=3.5in, clip=true, trim=0in 0in 0in 0in]{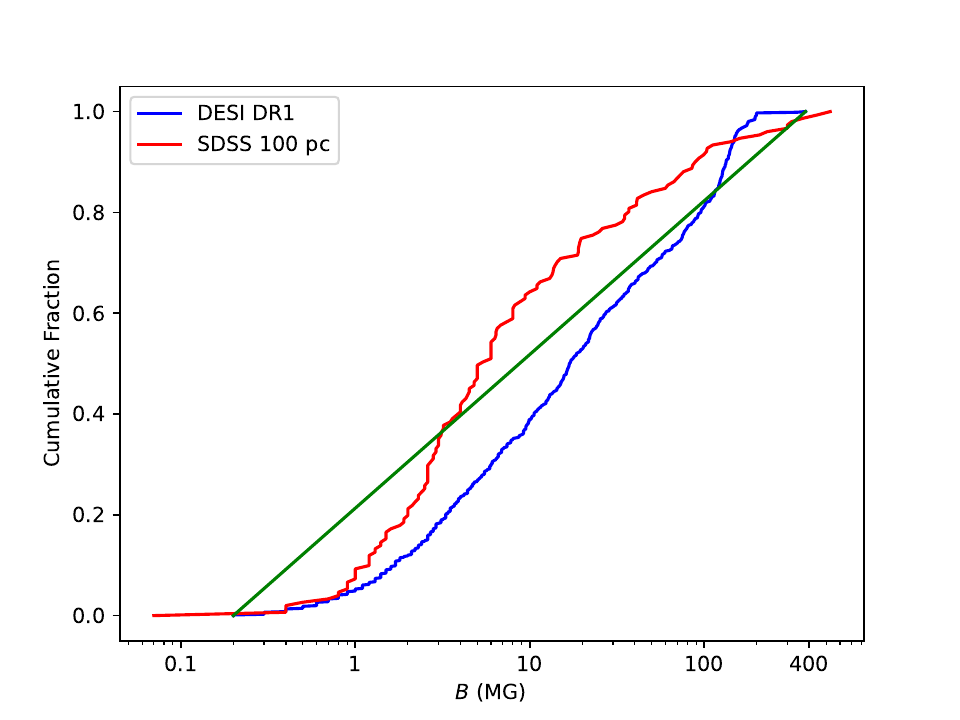}
    \caption{Top: Field strength distribution from our fits to the MWDS in DESI DR1 and the SDSS 100 pc sample \citep{Moss25b}. The DESI sample contains many more objects, especially those at higher field strengths. Both surveys are biased against field strengths below 0.1 MG. Bottom: Cumulative distribution of field strengths from the DESI DR1 sample (blue) and SDSS 100 pc sample (red)  compared to a log-uniform distribution (green). The DESI sample shows an excess of high field objects and a lack of weak field objects compared to the SDSS 100 pc sample and the log-uniform distribution.}
    \label{fig9} 
\end{figure}

Figure \ref{fig9} shows the distribution of field strengths for the MWDs with model fits in our DESI DR1 sample compared to the SDSS 100 pc sample \citep{Moss25b}. While many objects have weaker fields strengths ($B < 10$ MG), stronger fields are fairly common, particularly those above 100 MG. It is worth noting that above 100 MG, absorption features tend to be smeared out, which creates degeneracies between the geometry and field strength. We classify these objects as DAH? in our identification, as we cannot determine the atmospheric composition without absorption features. Hence the true field strength for these objects cannot be precisely constrained. However, the field strengths for these objects must be on the order of 100 MG, such that the exact values do not impact our results. 

Our sample contains many more high field objects compared to the SDSS 100 pc sample. Many of these high-field objects are young and are low to intermediate mass, such that they are intrinsically brighter and are thus overrepresented in magnitude-limited samples. \citet{Bagnulo21} found a roughly constant field strength distribution per dex in the 20 pc sample. We instead find a lack of MWDs with field strengths below $\sim$10 MG and an excess of objects above $\sim$100 MG compared to the log-uniform distribution. 

\begin{figure}[!ht]
    \centering
    \includegraphics[width=3.5in, clip=true, trim=0in 0in 0in 0in]{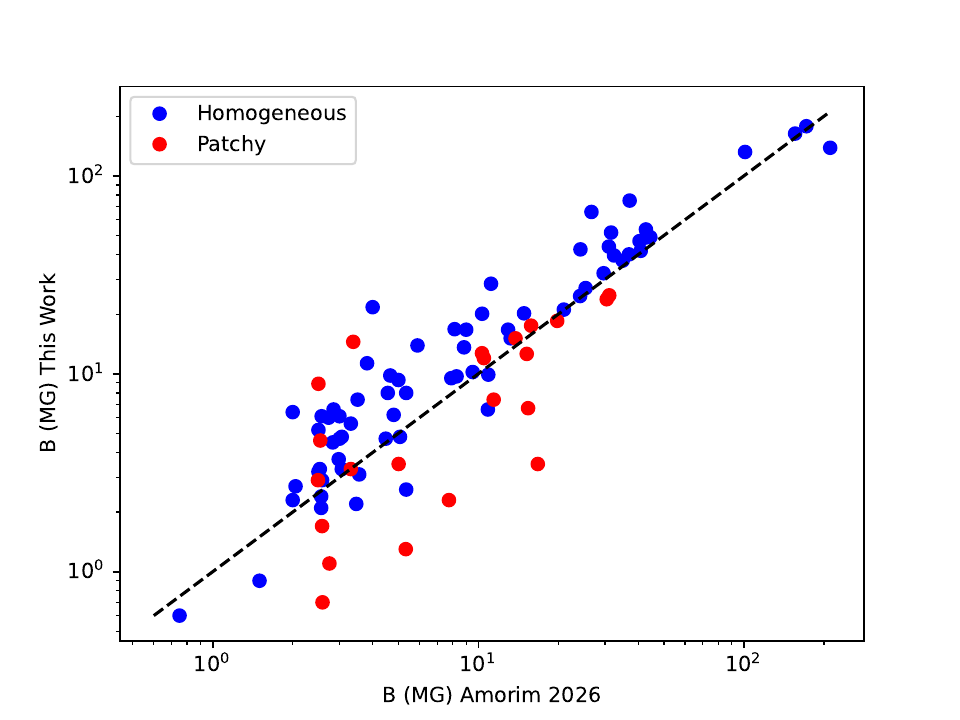}
    \caption{Comparison between the field strengths derived from \citet{Amorim26a} and this work. The estimates agree within reason across a range of field strengths, though our objects with patchy fits often return notably weaker field estimates.}
    \label{figlarissa} 
\end{figure}

\citet{Amorim26a} identified 137 new MWDs in DESI DR1 and obtained B-field estimates for these objects, 91 of which are in our sample and have magnetic fits. Figure \ref{figlarissa} shows the comparison of our field strengths for these objects. There is reasonable agreement between the values, particularly for the objects that we use a homogeneous model for. However, many of the objects that we fit with a patchy model tend to have weaker field estimates than what \citet{Amorim26a} finds. As we showed in the case of WDJ044543.74$-$031346.97, the patchy model often returns weaker field strengths than the homogeneous model, particularly for cooler objects that show weak line profiles.

\begin{figure*}[!ht]
    \centering
    \includegraphics[width=8in, clip=true, trim=0in 0in 0in 0in]{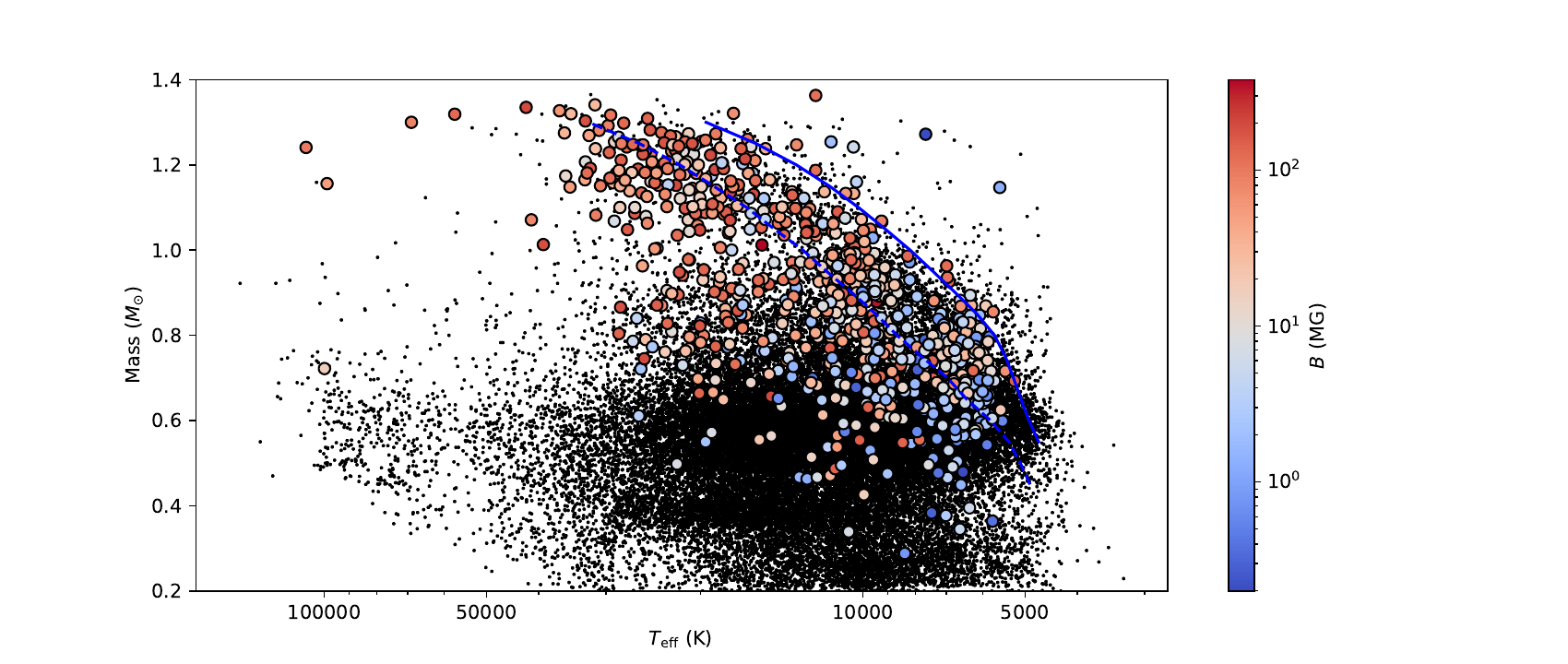}
    \caption{Photometric masses as a function of effective temperature with MWDs colored by their field strength. Objects with weaker fields are much more common at lower masses and cooler temperatures. The dashed and solid lines are the onset of and 80\% crystallization tracks respectively.}
    \label{fig10} 
\end{figure*}

Figure \ref{fig10} shows the stellar parameters of the sample but with MWDs colored by field strength. Objects with weak fields are significantly more common among low mass and colder objects, a trend well-studied in the literature \citep{Bagnulo22,Moss25b}. Higher mass objects likely form via mergers, which can reset their cooling age and produce stronger fields, while average mass objects likely come from single-star evolution, yielding weaker fields that take several Gyrs to reach the surface.  

\begin{figure}[!ht]
    \centering
    \includegraphics[width=3.7in, clip=true, trim=0in 0in 0in 0.5in]{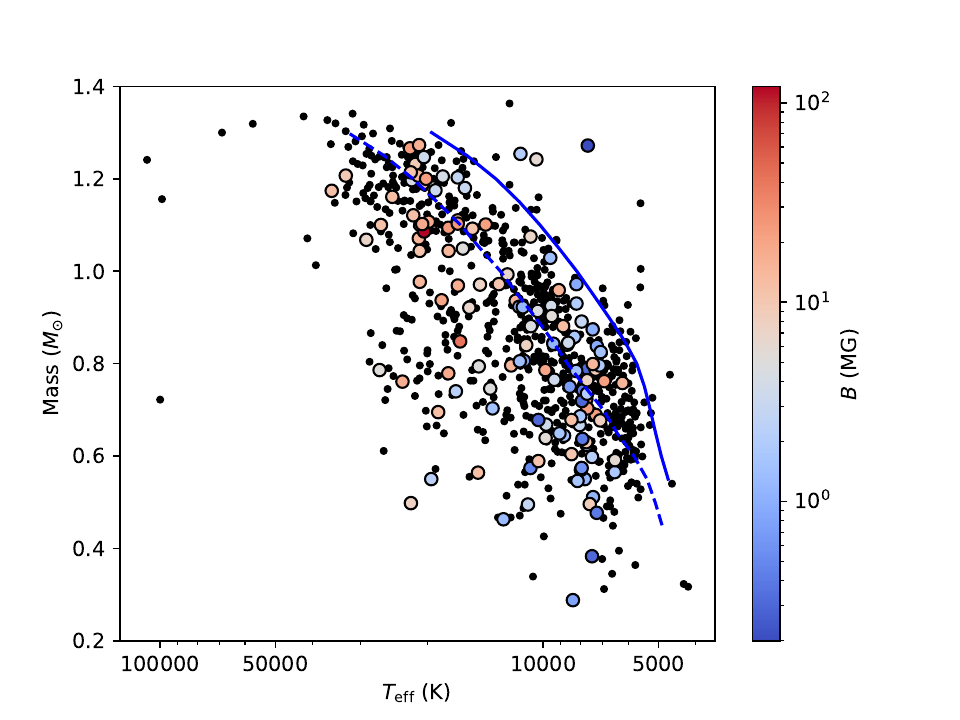}
    \caption{MWDs with homogeneous fits (black points) and patchy fits (colored points). The patchy objects occupy a similar parameter space as those that can be fit with the homogeneous model.}
    \label{figpatchy} 
\end{figure}

If the objects that require patchy fits share a mechanism to produce their fields and atmospheres, they may show trends in their stellar parameters that are distinct from objects fit with a homogeneous model. Figure \ref{figpatchy} shows the parameters of our patchy atmosphere objects compared to the rest of the MWDs in our sample. The patchy objects occupy a similar range of masses and effective temperatures as those that can be fit with the homogeneous model. Additionally, there is no discernible trend in field strength beyond the fact that cooler, lower mass objects have weaker fields, but this is true among all the MWDs. Hence the objects that require patchy atmospheres do not appear to share anything in common besides their inability to be fit with a homogeneous model, so they may not share a common mechanism that produces their unique fields.

\begin{figure}[!ht]
    \centering
    \includegraphics[width=3.5in, clip=true, trim=0in 0in 0in 0in]{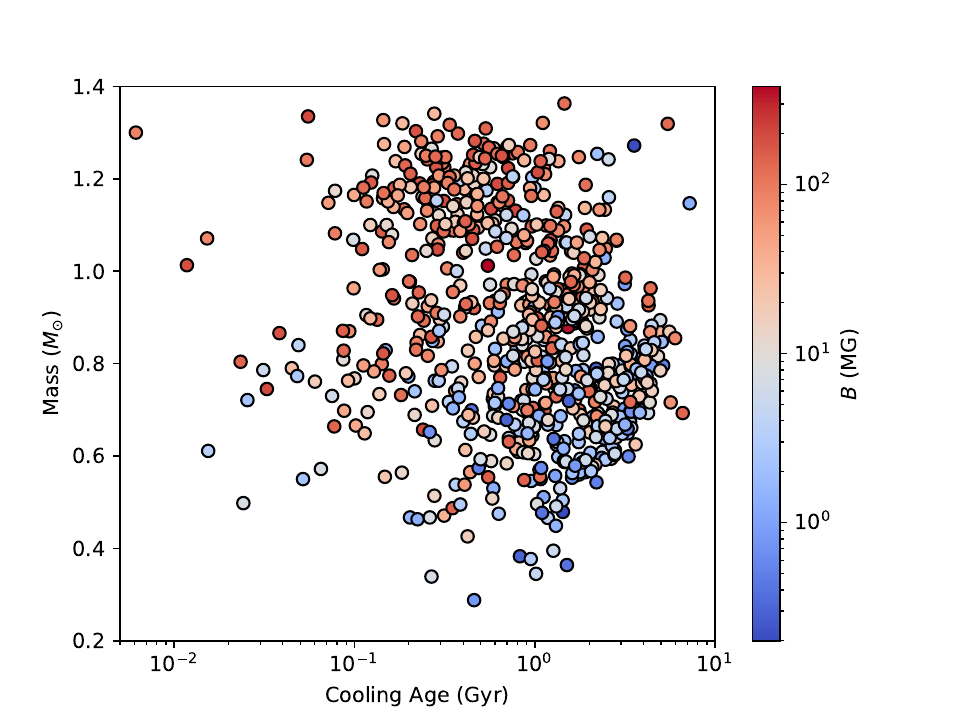}
    \caption{Mass as a function of cooling age for the DESI DR1 MWDs. Points are colored based on the best fit field strength.}
    \label{fig11} 
\end{figure}

Given the nonlinear cooling of WDs, it is more useful to analyze trends in the MWDs by looking at their cooling ages rather than their effective temperatures. Figure \ref{fig11} shows the masses as a function of cooling age for the MWDs colored by field strength. Objects older than $\sim$1.5 Gyr and between 0.6 and 0.9 $M_{\odot}$ typically have fields weaker than 10 MG, while stronger fields are more prevalent among younger objects regardless of mass.   

\subsection{Gaussian Mixture Model Analysis}

Given the dual formation channel of MWDs, \citet{Moss25b} used a Gaussian Mixture Model \citep[GMM]{Day69,Mclachlan00} to estimate the locations of individual clusters in the mass-cooling age-field strength parameter space. We employ a similar method here using \texttt{scikit-learn}'s \texttt{GaussianMixture} model \citep{scikit} which calculates the optimal means and covariance matrices of the Gaussian distributions that describe our sample (see \citealt{Burrow20} for details). We make two adjustments however in this analysis. One is that we use $\log_{10}(B_d)$ in our calculation rather than the raw field strengths. As the range of field strengths spans multiple orders of magnitude (0.2$-$380.8 MG), the distribution of these values is intrinsically non-Gaussian and thus poorly approximated by the GMM in linear space. Second, we test which number of clusters is optimal by calculating the Bayesian Information Criterion \citep[BIC,][]{Schwarz78} and the Integrated Completed Likelihood \citep[ICL,][]{Biernacki00}. The BIC balances the model fit with complexity by penalizing the addition of free parameters and is defined as: 

\begin{equation}
    BIC = k\ln(n)-2\ln(L)
\end{equation}

\noindent where $L$ is the model's maximized likelihood, $k$ is the number of free parameters, and $n$ is the sample size. A lower BIC yields an improved trade-off between fit and model simplicity, as more clusters will almost always fit the data better but introduces more means and covariances. The ICL incorporates the BIC but takes into account how well each point is described by its assigned cluster and is defined as:

\begin{equation}
    ICL = BIC +2H
\end{equation}

\noindent where $H$ is the entropy of the cluster membership probabilities and is calculated by summing the probabilities $p$ over every data point $i$ and every cluster $j$:

\begin{equation}
    H=(-\Sigma{p_{ij}})(\ln{p_{ij}})
\end{equation}

\noindent More separation between clusters, and thus less ambiguous cluster membership probabilities, yields a lower ICL. We calculate the BIC and ICL values for $N=1,2,...6$ clusters. 

\begin{figure*}[!ht]
    \centering
    \includegraphics[width=3.5in, clip=true, trim=0in 0in 0in 0in]{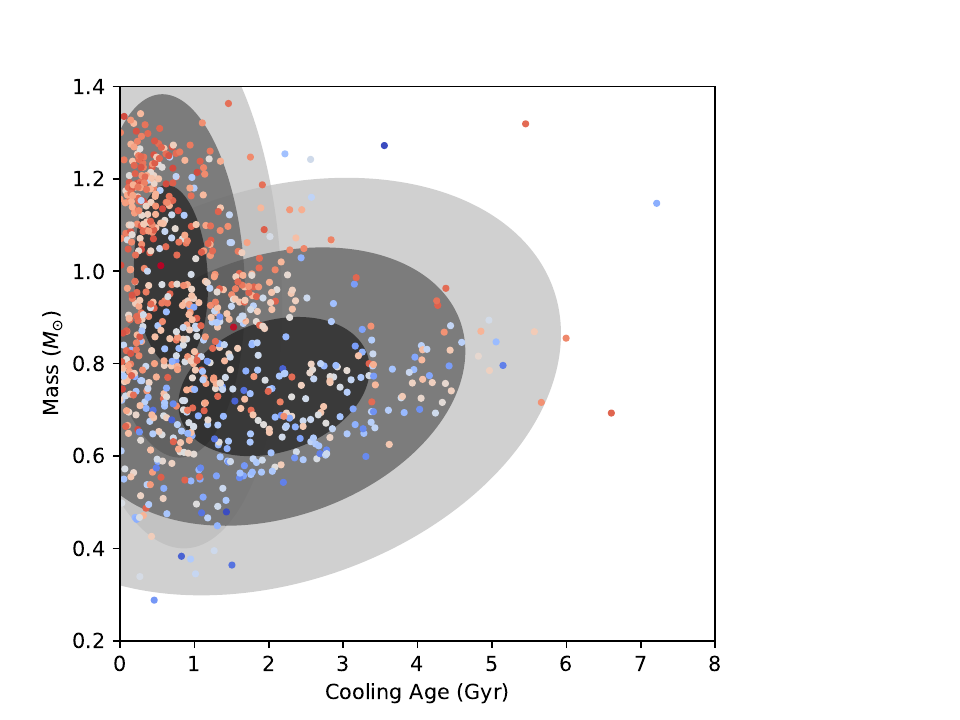}
    \hspace{-0.8in}
    \includegraphics[width=3.5in, clip=true, trim=0in 0in 0in 0in]{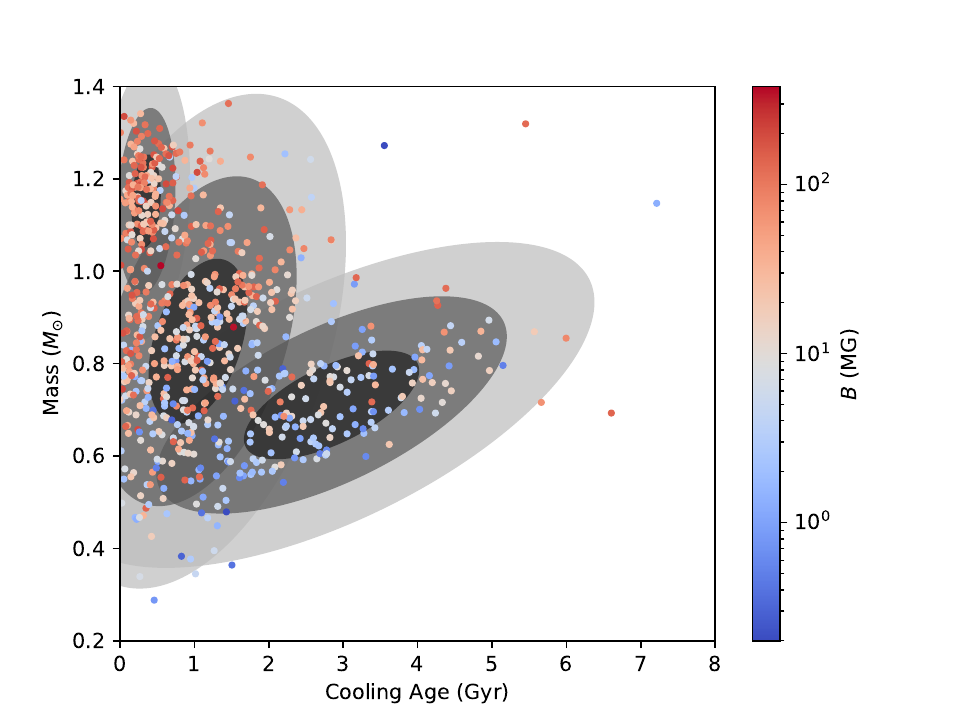}
    \caption{Our GMM analysis using the masses, cooling ages, and $\log_{10}(B_d)$ as inputs. Here we show the analysis with 2 (left panel) and 3 (right panel) clusters. The shaded regions represent the 1, 2, and 3$\sigma$ intervals from dark to light gray respectively. In both cases there is distinct separation between each cluster at the $1\sigma$ level.}
    \label{figgmm} 
\end{figure*}

Figure \ref{figgmm} shows our GMM calculation using 2 and 3 clusters. We obtain distinct separation between the clusters at the 1$\sigma$ level in both cases. However, 3 clusters minimizes the ICL while 5 clusters minimizes the BIC. Since minimizing the BIC is possible typically by simply adding more clusters, the ICL is a more valuable diagnosis. In addition, the 3 cluster model captures important details about the sample: the lowest mass cluster is well-centered on the old, lower-mass MWDs that likely come from single-star evolution, while the highest mass cluster is well-centered among the ultramassive MWDs which likely have significant contributions from mergers. Meanwhile, the third cluster captures the intermediate mass range that contains many young objects though not quite as young as the ultramassive ones, and also contains many objects with weaker fields than the ultramassive population. We further discuss the origins of objects in these clusters in Section 4.4. The high, intermediate, and low mass cluster centers occur at 0.35 Gyr, 1.14 $M_{\odot}$, 49.9 MG; 1.06 Gyr, 0.85 $M_{\odot}$, 16.6 MG; and 2.85 Gyr, 0.71 $M_{\odot}$, 4.75 MG respectively. 

\subsection{The Frequency of Magnetism}

\begin{figure}[!ht]
    \centering
    \includegraphics[width=3.5in, clip=true, trim=0in 0in 0in 0in]{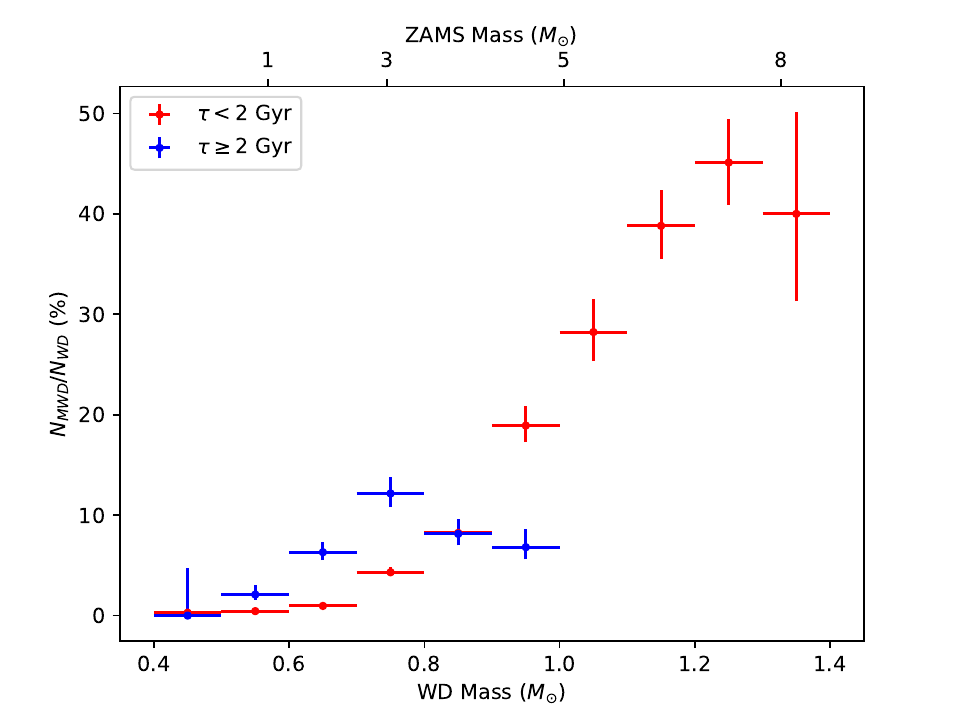}
    \includegraphics[width=3.5in, clip=true, trim=0in 0in 0in 0in]{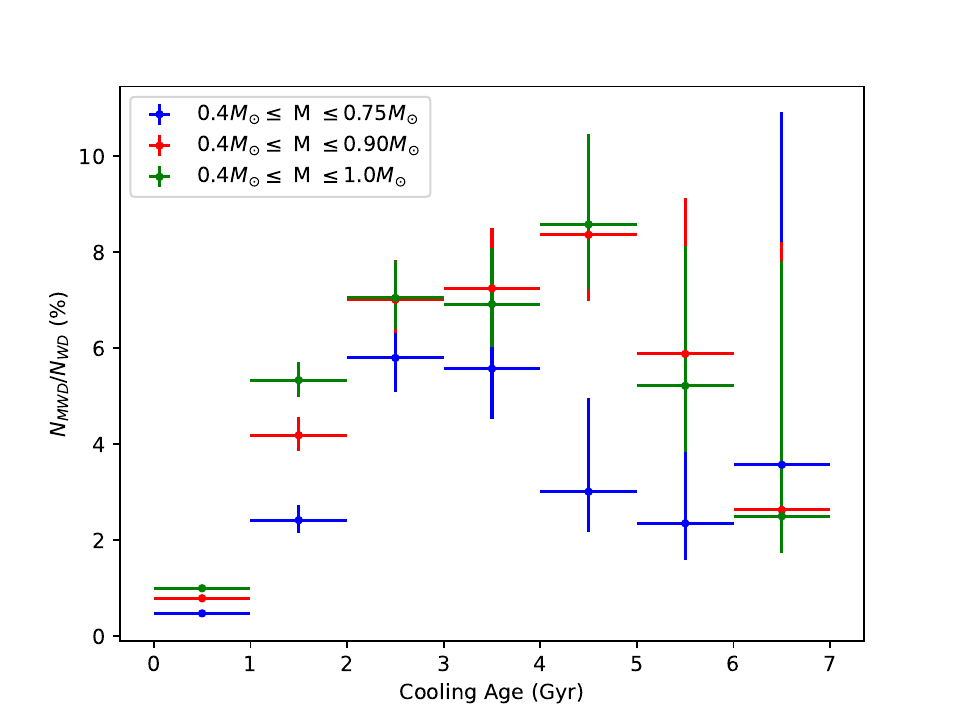}
    \caption{Magnetic fraction as a function of mass (top panel) and cooling age (bottom panel). We consider only objects with hydrogen-dominated atmospheres when determining the magnetic fraction. We also exclude objects below 0.4 $M_{\odot}$ as these are too low mass to have formed via single-star evolution. In the top panel we also exclude old objects above 1.0 $M_{\odot}$ due to small number statistics. The horizontal error bars represent the width of each bin. The mass bins are 0.1 $M_{\odot}$ wide while the age bins are 1 Gyr wide. The ZAMS masses in the top panel are calculated using the IFMR from \citet{Cummings18}.}
    \label{fig12} 
\end{figure}

Analyzing the fraction of MWDs relative to nonmagnetic objects across mass and cooling age can provide insight as to the origin of the fields. Figure \ref{fig12} shows the magnetic fraction as a function of mass and cooling age for the objects with hydrogen-dominated atmospheres. When calculating the fraction as a function of mass, we split the entire sample into two groups: WDs that are either younger or older than 2 Gyr. This is to distinguish between MWDs that likely form via mergers (young) and those that form via single-star evolution (old). Similarly, we use three separate mass ranges when calculating the fraction as a function of cooling age. The cutoffs for these bins are somewhat arbitrary but have similarly been used in analyzing volume-limited samples \citep{Bagnulo22,Moss25b}. We calculate the errors in the magnetic fraction using the binomial probability distribution discussed in \citet{Burgasser03}. 

Among older objects that likely form via single-star evolution, the magnetic fraction shows a clear peak in the $0.7-0.8$ $M_{\odot}$ bin similar to the SDSS 100 pc sample \citep{Moss25b}. Young objects show a clear increase in their magnetic fraction from 0.6 all the way up to 1.3 $M_{\odot}$. The key advantage of DESI is also on display here: the large number of objects leads to tighter error bars in general, making the trends noticeably clearer. 

The magnetic fraction as a function of cooling age shows similar trends to the 20 and 100 pc samples depending on the mass range considered. When considering objects less than 0.75 $M_{\odot}$, the fraction peaks between 2 and 4 Gyr as it does in the 100 pc sample \citep{Moss25b}. However, if higher mass WDs are included, the peak extends out to $4-5$ Gyr similar to the 20 pc sample \citep{Bagnulo21,Bagnulo22}, suggesting that it takes longer for fields to appear among higher mass MWDs that are less likely to come from binary evolution. 

A 0.75 $M_{\odot}$ WD corresponds to a $\sim$3 $M_{\odot}$ star on the main-sequence. Assuming these MWDs come from single-star evolution, stars at this mass appear to make MWDs more frequently than stars of lower and higher mass. The peak of this magnetic fraction should not be impacted by the biases of the DESI sample nor the biases of using Zeeman splitting to identify magnetic objects. At a given effective temperature, low mass WDs are brighter than high mass ones, such that we should have more low mass objects in our sample rather than fewer. Weaker fields do tend to be more common among lower mass objects than higher mass ones such that we could biased against those objects. However in the 20 pc sample, which utilized polarimetry to detect fields inaccessible to spectroscopy, there are six MWDs between 0.6 and 0.7 $M_{\odot}$ that are older than 2 Gyr and have hydrogen-dominated atmospheres \citep{Bagnulo21}. 4 of these are hot and have strong fields such that they can be detected via spectroscopy. For comparison, there are four between 0.7 and 0.8 $M_{\odot}$ and older than 2 Gyr, two of which can be detected via spectroscopy. Hence while we are biased against weak fields, it is not obvious that this bias would create the dropoff in MWDs going from 0.75 $M_{\odot}$ to 0.65 $M_{\odot}$. Similarly, we should not be more biased against detecting MWDs from $0.8-0.9$ $M_{\odot}$ as these should not possess fundamentally weaker fields than those between 0.7 and 0.8 $M_{\odot}$. Hence the peak in the magnetic fraction at 0.75 $M_{\odot}$ is a real feature of our sample.

\subsection{The Origins of Magnetism}

The large sample of MWDs combined with advancements in calculations of magnetic field evolution allow us to examine the origin of white dwarf magnetism at an in-depth level. Many of the high mass, strong field objects in the sample likely come from binary evolution, with a significant contribution from double WD mergers \citep{Berro12,Briggs15,Temmink20,Jewett24}. On the other hand, average mass white dwarfs likely come mostly from single-star evolution \citep{Temmink20}, and thus a magnetic field either originates from a prior evolution stage or crystallization. In the crystallization dynamo scenario, oxygen enrichment of the core generates convective flows, potentially spinning up a dynamo \citep{Isern17,Schreiber21,Ginzburg22}. However, this mechanism runs into several challenges. One is that the generated velocities from the chemical separation may not actually be fast enough to create a dynamo \citep{Montgomery24}. Additionally, it takes multiple Gyrs for a magnetic field to reach the surface after forming in the center of the white dwarf \citep{Blatman24a,Blatman24b}. Hence many MWDs are not old enough to have a detectable surface field if that field indeed came from crystallization. Lastly, while the incidence of magnetism often correlates with the onset of crystallization, many MWDs have not begun crystallizing as seen in Figure \ref{fig7}. Hence crystallization alone cannot be the sole source of magnetism among lower mass MWDs even if it is viable in some. 

\begin{figure*}[!ht]
    \centering
    \includegraphics[width=3.5in, clip=true, trim=0in 0in 0in 0in]{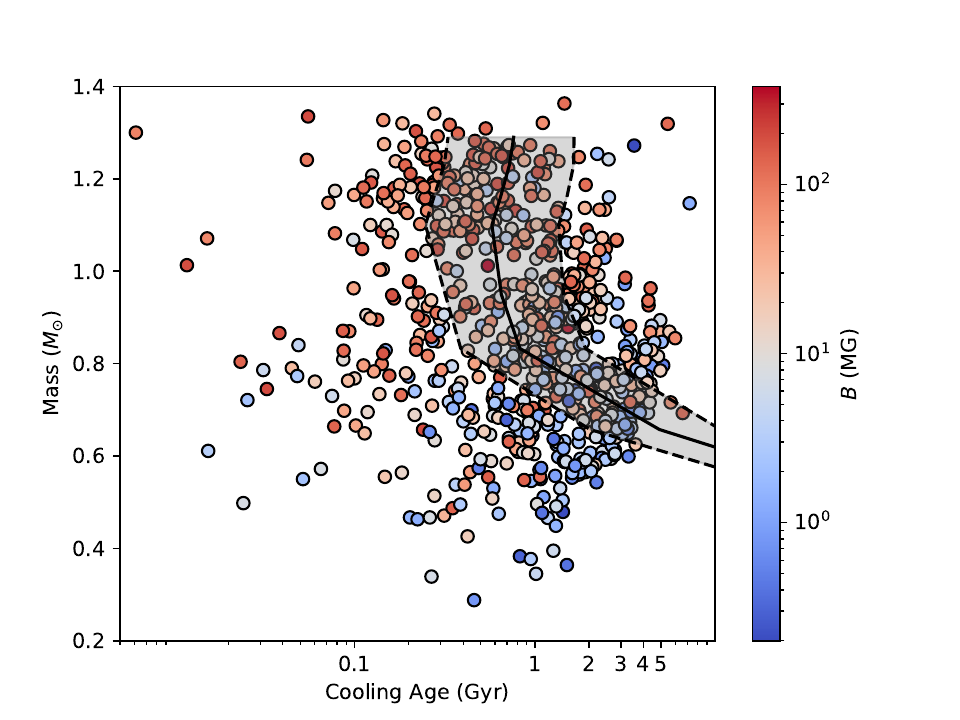}
    \includegraphics[width=3.5in, clip=true, trim=0in 0in 0in 0in]{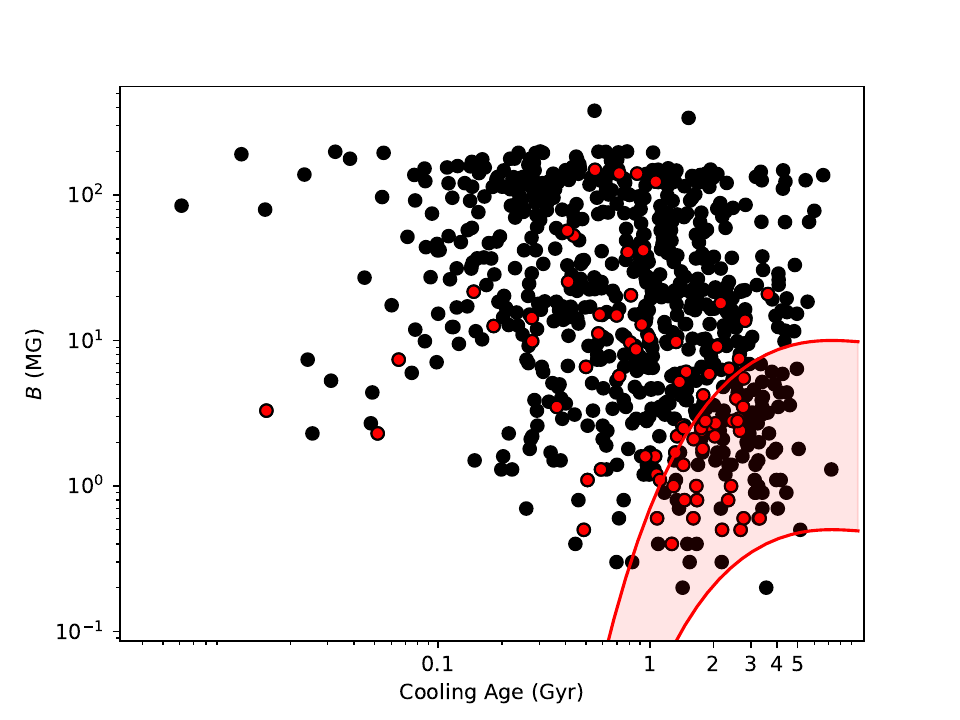}
    \caption{Left panel: Same as Figure \ref{fig11} but with the predicted breakout times for a core-convective dynamo that emerges during the white dwarf phases (solid black line) calculated by \citet{Camisassa24}. The dashed black lines represent an error of $\pm0.3$ dex. This dynamo can explain many of the MWDs in our sample but fails for those at very young ages and low masses. Right panel: Field strength as a function of cooling age for our DESI MWD sample with objects between 0.5 and 0.64 $M_{\odot}$ highlighted in red. The red region corresponds to the predicted field strengths and cooling ages for a dynamo that fills the radiative interior on the red giant phase as calculated by \citet{Einramhof26}. Many of the low mass MWDs can be explained by this dynamo. However, like the core-convective dynamo, many of the MWDs in our sample are too young to be explained by this formation channel.}
    \label{figcore} 
\end{figure*}

In contrast, recent calculations have shown that a magnetic field that forms on the main-sequence and is confined to the convective core throughout the giant phase can reproduce the cooling ages and field strengths of many average mass MWDs in volume-limited samples \citep{Camisassa24,Castro26}. The left panel of Figure \ref{figcore} shows the predicted breakout times for a core-convective dynamo that emerges at the white dwarf surface overlaid with our DESI MWD sample. Similar to the SDSS 100 pc sample, the track runs almost cleanly through the middle of the mass-cooling age distribution \citep{Moss25b}. Both average and higher mass MWDs lie within the $\pm1\sigma$ errors of the predicted breakout time, such that this core-convective dynamo could explain some of the higher mass MWDs that do not necessarily come from binary evolution. 

One limitation of the core-convective dynamo is that it can only explain MWDs above 0.65 $M_{\odot}$ \citep{Camisassa24}. The breakout time for MWDs below this limit is on the order of the Hubble time under this mechanism, hence the core-convective dynamo cannot explain these objects. An altered version of this scenario is if the core dynamo is able to leak out of the core and fill the radiative interior of the star as it evolves on to the red giant branch. \citet{Einramhof26} calculated the evolution of this field for a 1.5 $M_{\odot}$ using the measurements of fields within red giant interiors to calibrate their field strengths. The right panel of Figure \ref{figcore} shows the predicted field strengths at the surface of the white dwarf as a function of cooling age under this dynamo mechanism overlaid with our DESI MWD sample. This mechanism specifically operates for MWDs between 0.5 and 0.64 $M_{\odot}$, so we highlight objects between those masses in red. Many of these lower mass MWDs can be explained by this dynamo, though there are still many at younger ages and higher field strengths than what this dynamo predicts. These latter objects could be the products of binary mergers such as two red giants, a red giant and a main-sequence star \citep{Berro12,Briggs15,Temmink20} etc.

The distribution of MWDs in this sample is best modeled with 3 clusters using a GMM. While the parameters of objects in the high and low mass clusters can be explained by the relevance of binary evolution in each cluster, the intermediate cluster is more complicated. There is clearly a mixture of high and weak field objects in this cluster, with no obvious preference for one or the other. Hence disentangling the formation mechanisms of these intermediate objects is more challenging.

\begin{figure*}[!ht]
    \centering
    \includegraphics[width=3.5in, clip=true, trim=0in 0in 0in 0in]{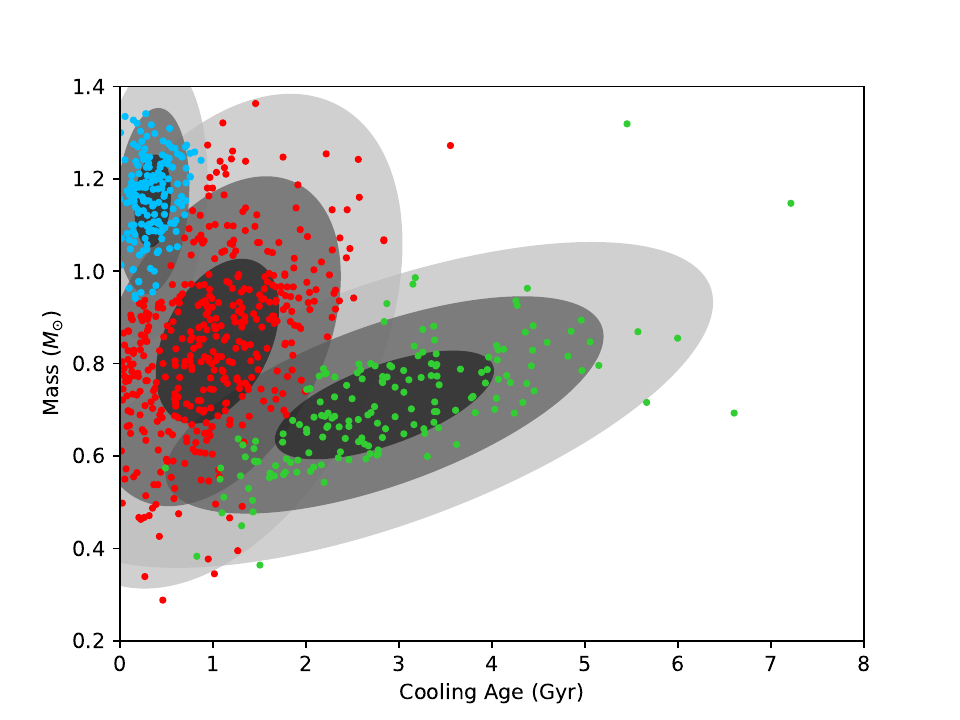}
    \includegraphics[width=3.5in, clip=true, trim=0in 0in 0in 0in]{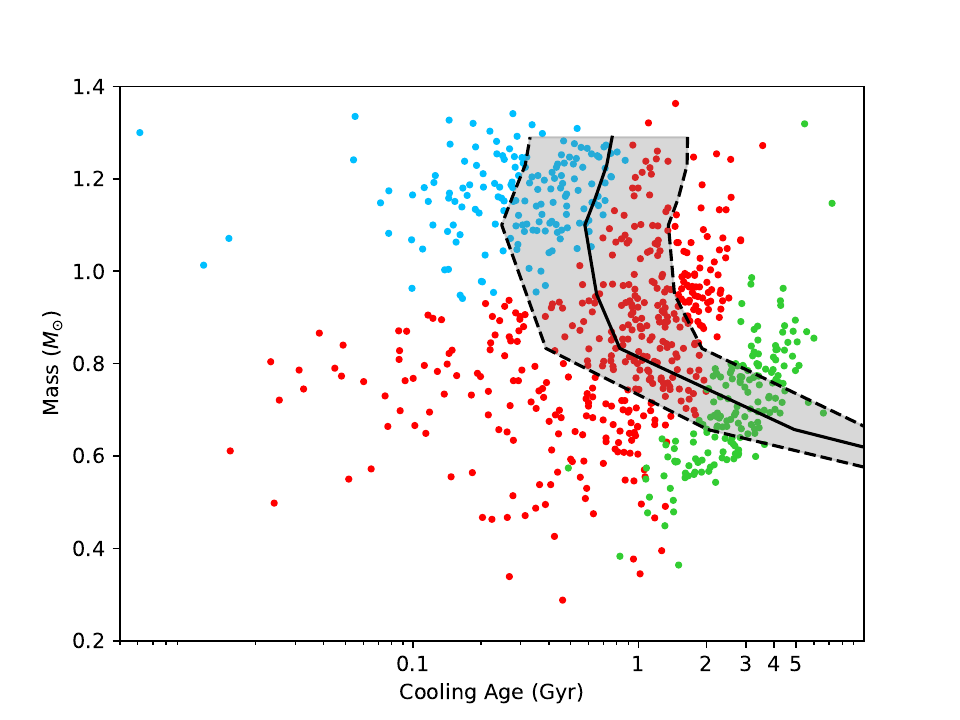}
    \caption{Left panel: Clusters calculated by the GMM and the masses and cooling ages of each MWD colored by which cluster each object is assigned to. Right panel: The predicted breakout times for a core-convective dynamo with the MWDs colored by their cluster assignment. Objects in the high and intermediate mass clusters (red and green, respectively) are frequently younger than the predicted dynamo timescale, indicating likely contributions from binary evolution.}
    \label{figpredict} 
\end{figure*}

One way to address this is to see which clusters the GMM assigns individual objects to. The GMM calculates the probability of a point belonging to a given cluster, and assigns that point based on which probability is highest. The left panel of Figure \ref{figpredict} shows which points are assigned to each cluster by the GMM. Since the clusters are widely separated at the 1$\sigma$ level, there is clean separation in the parameters of which points belong to which cluster. The right panel of Figure \ref{figpredict} shows how the member assignment compares to the core-convective dynamo. While many objects in the intermediate cluster can be explained by this dynamo, many others are younger or  lower mass than what this dynamo predicts. Additionally, these objects are too hot to have begun crystallization. Thus many of these young, intermediate mass objects likely came from mergers but not necessarily double white dwarf mergers given their lower masses \citep{Briggs15,Temmink20}.

Isolating which high mass objects likely come from mergers or single-star evolution is challenging without additional information such as rotation periods, spatial velocities, or unusual composition. However, even these do not necessarily tell the full picture. Only a few percent of MWDs typically show photometric variability from volume-limited samples \citep{Farihi24,Jewett24,Moss25b}, and there are biases that complicate identifying variable objects. If the magnetic and rotation axes are mostly aligned, variability will not be detected, and many high mass MWDs do not show variability in their polarized spectra \citep{Bagnulo24}. Additionally, some MWDs only show variability in specific bandpasses \citep{Rouis26}. Hence confirming rapid rotation in these objects remains a challenge. Furthermore, many likely merger remnants that are high mass, rapidly rotating, and strongly magnetic do not necessarily stick out in their kinematics \citep{Jewett24,Moss25b}. Ultimately, a strong field is not enough by itself to confirm a merger origin for a massive white dwarf, \and likely merger remnants do not necessarily possess magnetic fields (e.g. \citealt{Rouis26}). Hence the entire parameter space must be considered when determining the origin of these objects.

The young, average mass MWDs that appear in this sample are likely due to the bias towards brighter objects in DESI compared to volume-limited samples. There are only five MWDs younger than 1 Gyr and less than 0.8 $M_{\odot}$ in the SDSS 100 pc sample \citep{Moss25b}, and only one in the 20 pc sample \citep{Bagnulo21}. These objects almost certainly come from some form of binary evolution, as they are too young to have begun crystallizing or for their fields to have reached the surface under the aforementioned stellar dynamo scenarios. 

\section{Conclusions}

We present a detailed model atmosphere analysis of the MWDs in DESI DR1. We identified 831 MWDs from DESI DR1, which is the largest single data release of MWDs to date. We obtained spectral fits for 760 of these objects using an offset dipole model to determine the  magnetic parameters. Of these 760 objects, 126 possess either very shallow or very sharp absorption features, such that the homogeneous offset dipole model fails to reproduce the spectrum. For these, we employ a patchy offset dipole model to obtain significantly improved fits. We also identify 37 objects with Balmer emission features \citep{Manser23}, and obtain excellent fits to two of them using the offset dipole model. 

Compared to volume-limited samples, we find many hot MWDs that are too young to have begun core crystallization. Given the long breakout times of a crystallization-induced dynamo to reach the surface, it is unlikely that crystallization can explain many of the magnetic objects in our sample. In contrast, a stellar dynamo matches the parameters of many more MWDs across a wide range of masses, cooling ages, and field strengths. Many of the most massive and strongly magnetic targets however are likely merger remnants. In addition, the young MWDs that are more frequent in magnitude-limited samples such as DESI must have undergone some form of binary evolution, as they are too young and low mass for their fields to be generated via a stellar dynamo origin or crystallization. Thus a mix of origin sources is clearly necessary to produce magnetism in white dwarfs.

We determine the magnetic fraction as a function of mass and find that among younger objects, there is a clear increase in the fraction from 0.6 up to 1.3 $M_{\odot}$, confirming the likely merger remnants among the youngest MWDs. Among older objects, there is a peak in the magnetic fraction among likely single-star evolution products from $0.7 - 0.8$ $M_{\odot}$ similar to the SDSS 100 pc sample \citep{Moss25b}. This white dwarf mass corresponds to a progenitor mass of $\sim$3 $M_\odot$, indicating that stars of this mass could produce MWDs with fields detectable via spectroscopy more efficiently. We also calculate the magnetic fraction as a function of cooling age and find that the peak depends on what mass range we choose. For objects between 0.4 and 0.75 $M_{\odot}$, the peak occurs between 2 and 4 Gyr similar to the SDSS 100 pc sample. If we increase the upper mass limit to 0.9 or 1.0 $M_{\odot}$ however, the peak shifts to $4-5$ Gyr. Given the masses of these intermediate mass MWDs, it is difficult to conclusively determine if they come from binary evolution or single-star evolution. 

The sheer number of MWDs identified in this first DESI data release shows the impact large multiplex surveys can have on studying stellar magnetism. Future data releases will identify even more new objects, and will highlight variable systems by obtaining multi-epoch spectra. This will open up new avenues of exploration, particularly for studying the field structure of these objects, which could shed light into our targets that require patchy atmosphere fits. Given the multiple channels that produce MWDs, future analysis will need to analyze specific subsets of parameter space to truly determine the physics that govern each mechanism. With spectra on so many targets, this mission is now possible thanks to these multiplex surveys. The vast amounts of data, made evident by the analysis of the DESI DR1 white dwarfs, also highlights the need for advanced methods of identifying and classifying white dwarf spectra \citep{Vincent23,Vincent24,Munday26} rather than relying on visual inspection, though this remains challenging for MWDs given the nature of their spectral features.

\section{Use of AI-assisted technologies}
A large language model assistant, Claude, was used in preparation of this manuscript to aid in developing the topology maps in Figures \ref{figj0445map} and \ref{figj0056map}. The authors take full responsibility for the content of this paper.

\section{Acknowledgments}
This work is supported in part by the NSF under grant AST-2508429, the NASA under grants 80NSSC24K0380, 80NSSC24K0436, and 80NSSC22K0812, and the Smithsonian Institution.

Financial support for this publication comes from Cottrell Scholar Award \# CS-CSA-2026-091 from Research Corporation for Science Advancement.

Observations reported here were obtained at the MMT Observatory, a joint facility of the Smithsonian Institution and the University of Arizona.

This research used data obtained with the Dark Energy Spectroscopic Instrument (DESI). DESI construction and operations is managed by the Lawrence Berkeley National Laboratory. This material is based upon work supported by the U.S. Department of Energy, Office of Science, Office of High-Energy Physics, under Contract No. DE–AC02–05CH11231, and by the National Energy Research Scientific Computing Center, a DOE Office of Science User Facility under the same contract. Additional support for DESI was provided by the U.S. National Science Foundation (NSF), Division of Astronomical Sciences under Contract No. AST-0950945 to the NSF's National Optical-Infrared Astronomy Research Laboratory; the Science and Technology Facilities Council of the United Kingdom; the Gordon and Betty Moore Foundation; the Heising-Simons Foundation; the French Alternative Energies and Atomic Energy Commission (CEA); the National Council of Humanities, Science and Technology of Mexico (CONAHCYT); the Ministry of Science and Innovation of Spain (MICINN), and by the DESI Member Institutions: www.desi.lbl.gov/collaborating-institutions. The DESI collaboration is honored to be permitted to conduct scientific research on I'oligam Du'ag (Kitt Peak), a mountain with particular significance to the Tohono O'odham Nation. Any opinions, findings, and conclusions or recommendations expressed in this material are those of the author(s) and do not necessarily reflect the views of the U.S. National Science Foundation, the U.S. Department of Energy, or any of the listed funding agencies. 

\textit{Facilities:} Mayall (DESI)

\bibliography{DESIMWDs}{}
\bibliographystyle{aasjournalv7}

\end{document}